\pdfoutput=1

\documentclass[twocolumn,prc,aps,showkeys,superscriptaddress,nofootinbib,floatfix]{revtex4}

\usepackage{amssymb}
\usepackage{amsmath}
\usepackage{graphicx} % Include figure files
\usepackage{dcolumn}  % Align table columns on decimal point
\usepackage{rotating} 
\usepackage{subfigure} 
\usepackage{comment}
\usepackage{breakurl}

\usepackage{float}    % To make bordet around figures
\floatstyle{boxed}

\begin{document}

\title{Scintillating crystals for the Neutral Particle Spectrometer in Hall C at JLab}

\newcommand*{\CUA}{The Catholic University of America, Washington, DC 20064}

\newcommand*{\ANSL}{A.~I.~Alikhanyan National Science Laboratory, 
Yerevan 0036, Armenia}

\newcommand*{\JLAB}{Thomas Jefferson National 
Accelerator Facility, Newport News, Virginia 23606, USA}

\newcommand*{\IPNO}{Institut de physicque nucleaire d'Orsay, 15 rue Georges Clemenceau, 91406 Orsay, France}

\newcommand*{\SNU}{Seoul National University, 1 Gwanak-ro, Gwanak-gu, 08826 Seoul, Korea}

\newcommand*{\LCP}{Laboratoire de Chimie Physique, CNRS/Universit\'e Paris-Sud, B\^at. 349, 91405 Orsay, France}

\author{T.~Horn} 
\affiliation{\CUA}
\affiliation{\JLAB}

\author{V.V.~Berdnikov} 
\affiliation{\CUA}

\author{S.~Ali} 
\affiliation{\CUA}

\author{A.~Asaturyan}
\affiliation{\ANSL}

\author{M.~Carmignotto} 
\affiliation{\JLAB}

\author{J.~Crafts} 
\affiliation{\JLAB}

\author{A.~Demarque}
\affiliation{\LCP}

\author{R.~Ent}
\affiliation{\JLAB}

\author{G.~Hull} 
\affiliation{\IPNO}

\author{H-S.~Ko} 
\affiliation{\IPNO}
\affiliation{\SNU}

\author{M.~Mostafavi}
\affiliation{\LCP}

\author{C.~Munoz-Camacho} 
\affiliation{\IPNO}

\author{A.~Mkrtchyan}
\affiliation{\ANSL}

\author{H.~Mkrtchyan}
\affiliation{\ANSL}

\author{T.~Nguyen Trung} 
\affiliation{\IPNO}

\author{I.L.~Pegg}
\affiliation{\CUA}

\author{E.~Rindel} 
\affiliation{\IPNO}

\author{A.~Somov} 
\affiliation{\JLAB}

\author{V.~Tadevosyan} 
\affiliation{\ANSL}

\author{R.~Trotta} 
\affiliation{\CUA}

\author{S.~Zhamkochyan}
\affiliation{\ANSL}

\author{R.~Wang} 
\affiliation{\IPNO}

\author{S.~A.~Wood}
\affiliation{\JLAB}

\newpage
\date{\today}

%%%%%%%%%%%%%%%%%%%%%%%%%%%%%%%%%%%%%%%%%%%%%%%%%%%%%%%%

\begin{abstract}

This paper discusses the quality and performance of currently available PbWO$_4$ crystals of relevance to high-resolution electromagnetic calorimetry, e.g. detectors for the Neutral Particle Spectrometer at Jefferson Lab or those planned for the Electron-Ion Collider. Since the construction of the Compact Muon Solenoid (CMS) at the Large Hadron Collider (LHC) and early PANDA (The antiProton ANnihilations at DArmstadt) electromagnetic calorimeter (ECAL) the worldwide availability of high quality PbWO$_4$ production has changed dramatically. We report on our studies of crystal samples from SICCAS/China and CRYTUR/Czech Republic that were produced between 2014 and 2019. 

\end{abstract}

\keywords{Electromagnetic calorimeters, scintillator, crystal, glass, photo-luminescence, radiation damage, Electron-Ion Collider}
\maketitle

%\linenumbers

\section{Introduction}
\label{intro}

Gaining a quantitative description of the nature of strongly bound systems is of great importance for our understanding of the fundamental structure and origin of matter. Nowadays, the CEBAF at Jefferson Lab has become the world's most advanced particle accelerator for investigating the nucleus of the atom, the protons and neutrons making up the nucleus, and the quarks and gluons inside them. The 12-GeV beam will soon allow revolutionary access to a new representation of the proton's inner structure. In the past, our knowledge has been limited to one-dimensional spatial densities (form factors) and longitudinal momentum densities (parton distributions). This cannot describe the proton's true inner structure, as it will, for instance, be impossible to describe orbital angular momentum, an important aspect for nucleon spin, for which we need to be able to describe the correlation between the momentum and spatial coordinates. A three-dimensional description of the nucleon has been developed through the Generalized Parton Distributions (GPDs)~\cite{Diehl03,Radyushkin96,Goeke01,Belitsky05} and the Transverse Momentum-Dependent parton distributions (TMDs)~\cite{Mulders96,Bacchetta07,Anselmino11,Cahn78}. GPDs can be viewed as spatial densities at different values of the longitudinal momentum of the quark, and due to the space-momentum correlation information encoded in the GPDs, can link through the Ji sum rule~\cite{Ji97} to a parton's angular momentum. The TMDs are functions of both the longitudinal and transverse momentum of partons, and they offer a momentum tomography of the nucleon complementary to the spatial tomography of GPDs.

The two-arm combination of neutral-particle detection and a high-resolution magnetic spectrometer offers unique scientific capabilities to push the energy scale for studies of the transverse spatial and momentum structure of the nucleon through reactions with neutral particles requiring precision and high luminosity. It enables precision measurements of the deeply-virtual Compton scattering cross section at different beam energies to extract the real part of the Compton form factor without any assumptions. It allows measurements to push the energy scale of real Compton scattering, the process of choice to explore factorization in a whole class of wide-angle processes, and its extension to neutral pion photo-production. It further makes possible measurements of the basic semi-inclusive neutral-pion cross section in a kinematical region where the QCD factorization scheme is expected to hold, which is crucial to validate the foundation of this cornerstone of 3D transverse momentum imaging. 

The Neutral-Particle Spectrometer (NPS) in Hall C will allow accurate access to measurements of hard exclusive (the recoiling proton stays intact in the energetic electron-quark scattering process) and semi-inclusive (the energy loss of the electron-quark scattering process gets predominantly absorbed by a single pion or kaon) scattering processes. To extract the rich information on proton structure encoded in the GPD and TMD frameworks, it is of prime importance to show in accurate measurements, pushing the energy scales, that the scattering process is understood. Precision measurements of real photons or neutral-pions with the NPS offer unique advantages here. 

The NPS science program currently features four fully approved experiments ~\cite{{E12-13-007},{E12-13-010},{E12-14-003},{E12-14-005}}.
E12-13-007 \cite{E12-13-007} will measure basic cross sections of the 
semi-inclusive $\pi^0$ electroproduction process off a proton target, at small 
transverse momentum (scale $P_{h\perp}\approx\Lambda$). These neutral-pion 
measurements will provide crucial input towards our validation of the basic 
SIDIS framework and data analysis at JLab energies, explicitly in terms
of validation of anticipated (x, z) factorization.
%In SIDIS $\pi^0$ electroproduction, the lack of diffractive $\rho$ contributions, the lack of pole contributions and thus radiative tail contributions at large z, the reduced nucleon resonance contribution (as for example compared to $ep\rightarrow e^{\prime}\pi^{-}\Delta^{++}$), and the proportionality to an average fragmentation function, are all points in favor to validate low-energy (x, z) factorization required to substantiate the SIDIS science output.
%
E12-13-010 will perform high 
precision measurements of the Exclusive Deeply Virtual Compton Scattering 
(DVCS) and $\pi^0$ cross section~\cite{E12-13-010}. 
%A wide range of kinematics accessible with an 11 GeV electron beam off an unpolarized proton target will be covered. 
The azimuthal, energy and helicity dependences of the cross section will all be
 exploited in order to separate the DVCS-BH interference and DVCS contributions
 to each of the Fourier moments of the cross section~\cite{Ji-1997}. 
%For each term, its Q$^2$ dependence will be measured independently. At the same time, the exclusive $\pi^0$ electroproduction cross section will also be measured and a longitudinal/transverse separation will be performed.
%
The goal of E12-14-003~\cite{E12-14-003} is to measure the 
cross-section for Real Compton Scattering (RCS) from the proton in Hall C at 
incident photon energies of 8 GeV (s = 15.9 GeV$^2$) and 10 GeV 
(s = 19.6 GeV$^2$) over a broad span of scattering angles in the wide-angle 
regime.
The precise cross-section measurements at the highest possible photon energies 
over a broad kinematic range will be essential in order to confirm whether the 
factorization regime has been attained and investigate the nature of the 
factorized reaction mechanism. 
%The recoil protons will be detected in the Hall C HMS magnetic spectrometer while the scattered photons in the new NPS photon spectrometer.
%
The differential cross section of the $\gamma p \rightarrow \pi^0 p$ process 
in the range of $10~GeV^2 < s <20~GeV^2$ at large pion center-of-mass angles of
 $55^o < \theta_{cm} < 105^o$ will be measured in experiment 
E12-14-005~\cite{E12-14-005}. 
Hard exclusive reactions provide an excellent opportunity to study the 
complicated hadronic dynamics of underlying subprocesses at partonic level. 
The exclusive photoproduction of mesons with large values of energy and momentum transfers ($ s \sim t \sim u >> \Lambda $) are among the most elementary reactions due to minimal total number of constituent partons involved in these $2 \rightarrow  2 $ reactions. 
%Existing world data on photoproduction of neutral pions on proton $ \gamma + p \rightarrow \pi^o + p $ have very large systematic errors and do not have suffcient accuracy to perform comprehensive phenomenological analysis. Here again the recoil proton will be detected in the HMS spectrometer while photons from the $\pi^o \rightarrow \gamma \gamma $ decay will be detected in the Neutral Particle Spectrometer (NPS).

\begin{figure}
\centering
\subfigure[\label{fig:NPS-right}] 
%caption for subfigure a
{\includegraphics[width=1.7in]{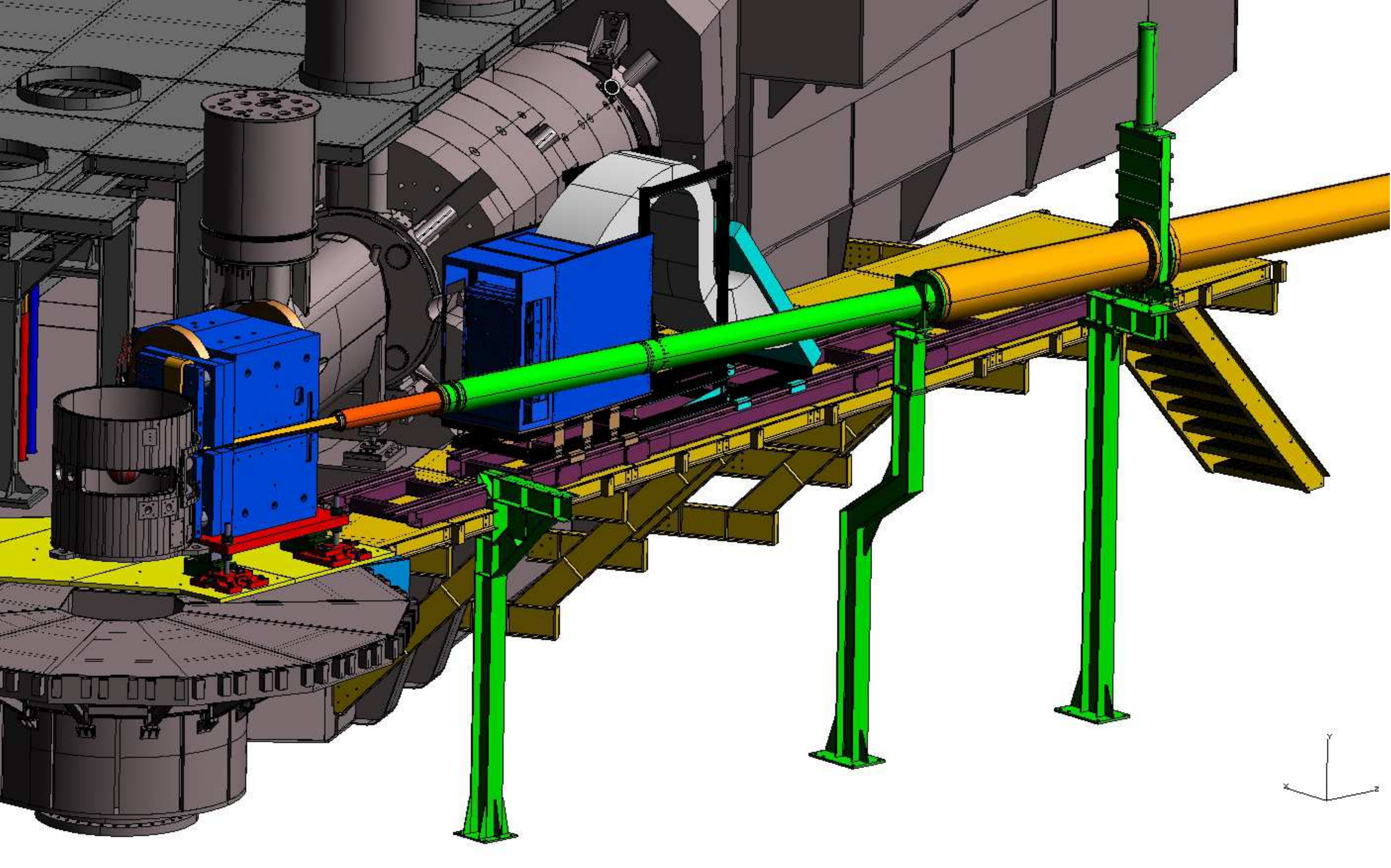} }
\hspace{0.1cm}
\subfigure[\label{fig:NPS-left}]
% caption for subfigure b
{\includegraphics[width=1.3in]{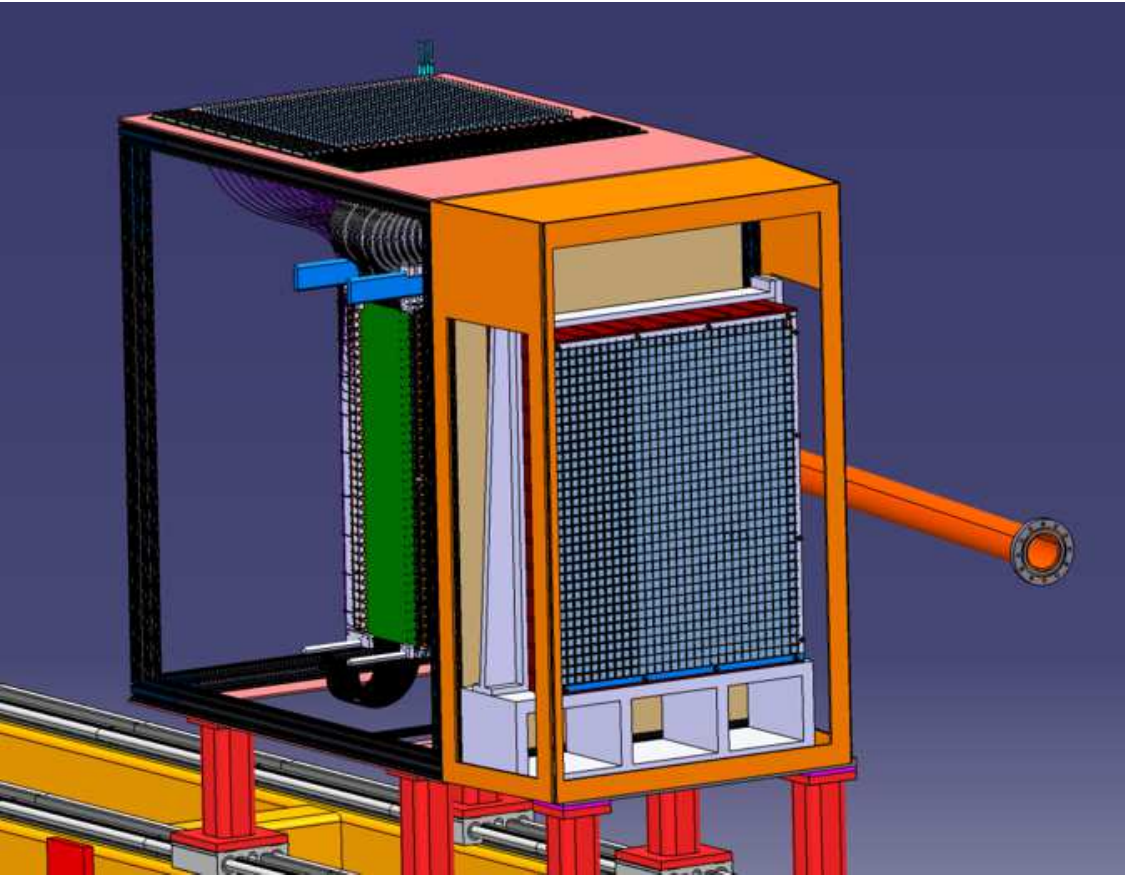} }
\caption{\label{fig:NPS-calorimeter} (Color online) Right: drawing of the NPS spectrometer in Hall C (right). The cylinder on lower left is the target, behind it in the pivot area is the NPS magnet, followed by the NPS calorimeter sitting on a rail system to allow for movement towards/away from the pivot. The dark gray structure is the SHMS; left: NPS calorimeter drawing with details of the crystal matrix inside the frame.}
\end{figure}

The NPS consists of an electromagnetic calorimeter preceded by a sweeping magnet. As operated in Hall C, it replaces one of the focusing spectrometers. To address the experimental requirements the NPS has the following components:
\begin{itemize}
\item{A 25 msr neutral particle detector consisting of ~1080 PbWO4 crystals in a temperature-controlled frame including gain monitoring and curing systems}
\item{HV distribution bases with built-in amplifiers for operation in a high-rate environment}
\item{Essentially deadtime-less digitizing electronics to independently sample the entire pulse form for each crystal}
\item{A vertical-bend sweeping magnet with integrated field strength of 0.3 Tm to suppress and eliminate charged background.}
\item{Cantilevered platforms off the Super-High Momentum Spectrometer (SHMS) carriage to allow for remote rotation. For NPS angles from 6 to 23 degrees, the platform will be on the left of the SHMS carriage (see Fig.~\ref{fig:NPS-right}); for NPS angles 23-57.5 degrees it will be on the right.}
\item{A beam pipe with as large opening/critical angle for the beam exiting the target/scattering chamber region as possible to reduce beamline-associated backgrounds}
\end{itemize}

Good optical quality and radiation hard PbWO$_4$ crystals are essential for the NPS calorimeter. Such crystals or more cost-effective alternatives are also of great interest for the Hall D Forward Calorimeter and the high-resolution inner calorimeters at the Electron-Ion Collider (EIC), a new experimental facility that will provide a versatile range of kinematics, beam polarizations and beam species, 
which is essential to precisely image the sea quarks and gluons in nucleons in 
nuclei and to explore the new QCD frontier of strong color fields in nuclei and
 to resolve outstanding questions in understanding nucleons and nuclei on the 
basis of QCD. One of the main goals of the EIC is the three-dimensional imaging of nucleon 
and nuclei and unveiling the role of orbital angular motion of sea quarks and 
gluons in forming the nucleon spin. Details about the EIC science, detector requirements, and
design considerations can be found in the EIC White Paper~\cite{Accardi} and Detector Handbooks~\cite{EIC}. 

The common requirements of these electromagnetic calorimeters on the active scintillating material are:
1) good resolution in angle to at least 0.02 rad  to distinguish between 
clusters, 
2) energy resolution to a few \%/$\sqrt{E}$ for measurements of the cluster 
energy, and 
3) the ability to withstand radiation down to at least 1 degree with respect to
 the beam line.
In this article we discuss the ongoing effort to understand the performance and selection of full-sized scintillator blocks for the NPS, as well as possible alternatives to crystals.

This article is organized as follows: section~\ref{sec-exp-req} describes the basic principle of neutral particle detection, specific NPS requirements, and specifications on the scintillator material, section~\ref{sec-crystals} reviews the scintillator fabrication, section~\ref{sec-quality-assur} describes experimental methods used in the investigation of the scintillator samples. The results of the measurements of scintillator properties, such as optical transmittance, emission spectra, decay times, light yield, and light yield uniformity are discussed in section~\ref{sec-characterization-results}. Section~\ref{sec-results-rad-damage} discusses the results on radiation damage and possible curing strategies. Scintillator structure and impurity analysis are presented in section~\ref{sec-structural-analysis}. Section~\ref{sec-prototype} discusses the design, construction, and commissioning of a single counter to test the scintillator performance, section~\ref{sec-glass} contains an overview of alternative scintillator material, and sectiob~\ref{sec-summary} presents the summary and conclusions.

\section{Experimental requirements on neutral particle detection}
\label{sec-exp-req}

Electromagnetic calorimeters are designed to measure the energy of a particle as it passes through the detector by stopping or absorbing most of the particles coming from a collision. The summed deposited energy is proportional and a good measure of the incident energy. An important requirements is thus the linearity of the scintillator material light response with the incident photon energy, i.e. the energy resolution. The segmentation of the calorimeter provides additional information and allows for discriminating single photons from, e.g., DVCS and two photons from $\pi^0$ decay, and electrons from pions. 

The NPS science program requires neutral particle detection over an angular range between 6 and 57.3 degrees at distances of between 3 meter and 11 meter~\footnote{the minimum NPS angle at 3m is 8.5 degrees, at 4m it is 6 degrees} from the experimental target, and with 2-3 mm spatial and 1-2\% energy resolution. Electron beam energies of 6.6, 8.8, and 11 GeV will be used. The individual NPS experiment requirements are listed in Table~\ref{tab:experiment-requir}.
                             
{\small                                                                  
\begin{table*}                                                             
\caption{\label{tab:experiment-requir} NPS experiment requirements. Electron beam energies of 6.6, 8.8, and 11 GeV will be used.}
{\centering  \begin{tabular}{|c|c|c|c|c|c|}
\hline
Parameter                    & E12-13-010             & E12-14-007    & E12-14-003           & E12-13-005  \\
\hline
Photon angl. res. (mrad)     &   0.5-0.75           & 0.5-0.75     & 1-2              & 1-2        \\
\hline
Energy res. (\%)             & (1-2)/$\sqrt{E}$ & (1-2)/$\sqrt(E)$ & 5/$\sqrt{E}$      & 5/$\sqrt{E}$ \\
\hline
Photon energies (GeV)        &   2.7-7.6        & 0.5-5.7      & 1.1-3.4                       & 1.1-3.4                 \\
\hline
Luminosity (cm$^{-2}sec^{-1}$) & $\sim10^{38}$    & $\sim10^{38}$ &   $\sim10^{38}$     &   $\sim10^{38}$    \\
\hline
Acceptance (msr)             &   60\%/25 msr            &      & 60\%/25 msr            & 10-60\%/25 msr      \\
\hline
Beam current ($\mu$A)        &    5-50      & 5-50       &  5-60, +6\% Cu     &  5-60, +6\% Cu     \\
\hline
Targets                      &  10cm LH2    & 10 cm LH2     & 10 cm LH2             & 10 cm LH2     \\
\hline
\end{tabular}\par}
\end{table*}
}
%\end{widetext}

The photon detection is the limiting factor of the experiments. Exclusivity of the reaction is ensured by the missing mass technique and the missing-mass resolution is dominated by the energy resolution of the calorimeter. The scintillator material should thus have properties to allow for an energy resolution of $1-2\%/\sqrt(E)$. 

The expected rates of the NPS experiments in the high luminosity Hall C range up to 1 MHz per module. The scintillator material response should thus be fast, and respond on the tens of nanosecond level.

Given the high luminosity and very forward angles required in the experiments, radiation hardness is also an essential factor when choosing the detector material. The anticipated doses depend on the experimental kinematics and are highest at the small forward angles. Based on background simulations dose rates of 1-5 kRad/hour are anticipated at the most forward angles. The integrated doses for E12-13-010 are 1.7 MRad at the center and 3.4 MRad at the edges of the calorimeter. The integrated doses for the other experiments are $<$ 500 kRad. The ideal scintillator material would be radiation hard up to these doses. The ideal material would also be independent of environmental factors like temperature.

\subsection{Choice of scintillator material}

The material of choice for the NPS calorimeter is rectangular PbWO$_4$ crystals of 2.05 by 2.05 cm$^2$ (each 20.0 cm long). The crystals are arranged in a 30 x 36 matrix, where the outer layers only have to catch the showers. This amounts to a total of 1080 PbWO$_4$ crystals. For NPS standard configurations, each
crystal covers 5 mrad and the expected angular resolution is 0.5-0.75 mrad, which is comparable with the resolution of the High Momentum Spectrometer (HMS), one of the well established Hall C spectrometers. The energy resolution of PbWO$_4$ was parameterized for the Primex experiment in Ref.~\cite{Hycal}. There, a matrix of 1152 PbWO$_4$ crystals was used with incident photons energies of 4.9-5.5 GeV. The resulting parameterization is $\sigma$/$E$=0.009$\bigoplus$0.025/$\sqrt{E}$$\bigoplus$0.010/$E$, where $E$ is the incident beam energy. A $\pi^0$ missing mass resolution of $\sim$1-2 MeV and production angle resolution of $\sim$3mrad were obtained. and is consistent with NPS experiment requirements. 

The emission of PbWO$_4$ includes up to three components, and increases with increasing wave length~\cite{Belsky}: $\tau_1 \sim$5 ns (73\%); $\tau_2 \sim$14 ns (23\%) for emission of $\lambda$ in the range of 400-550 nm; $\tau_3$ has a lifetime more than 100 ns, but it is only $\sim$4\% of the total intensity. 
The time resolution of the calorimeter based on PbWO$_4$ is thus sufficient to handle rates up to $\sim$1 MHz per block.

PbWO$_4$ crystals suffer radiation damage~\cite{Batarin,Fyodorov,Achenbach,Dormenev}, but optical properties can be recovered~\cite{Zhu95}. Studies at LHC suggest that the conservative dose limit for curing is 50 to a few 100 krad~\cite{Nagornaya,Zhu98}. If energy resolution is not a big issue, the limiting dose may be increased to a few MRad. The NPS includes a light monitoring and curing system to recover the crytal optical properties. These systems were tested with a prototye as discussed in section~\ref{sec-results-rad-damage}. The scintillation light output, decay time, and radiation resistance of PbWO$_4$ are temperature dependent~\cite{Leqcoq,Chao,Krutyak}, with the light yield increasing at low temperature, but decay time and radiation resistance decreasing with temperatures. The NPS design will thus be thermally isolated and be kept at constant temperature to within 0.1$^o$C to guarantee 0.5\% energy stability for absolute calibration and resolution. 

\subsection{Specifications on Scintillator Material}
\label{crystals}

The experimental requirements shown in Table~\ref{tab:experiment-requir} can be translated into specifications on the scintillator material, e.g. PbWO$_4$ crystals. Besides specifications related to dimension and optical properties, minimum limits on radiation hardness are also defined for scintillator material fabricated for operation in a high radiation environment like for the NPS or the EIC. Table \ref{tab:pwo-specification} lists the physical goals and specifications for NPS in comparison to those for EIC and other projects. 

%\begin{widetext}
{\small

\begin{table*}
\caption{\label{tab:pwo-specification} PbWO$_4$ crystal quality specifications for NPS, EIC, HyCAL/FCAL, CMS, and PANDA. The measurements to determine these properties are discussed in the text.}
{\centering  \begin{tabular}{|c|c|c|c|c|c|c|}
\hline
Parameter                    & Unit   & NPS & Hy(F)CAL & EIC &  CMS & PANDA         \\
\hline
Light Yield (LY) at RT      & pe/MeV & $\geq$15 & $\geq$9.5 & $\geq$15 & $\geq$8 & $\geq$16 \\
\hline
LY (100ms)/LY(1$\mu$s)       & \%     &  $\geq$90    &  $\geq$90 & $\geq$90  & $\geq$90  & $\geq$90  \\     
\hline
Longitudinal Transmission    &      &     & & &  &                 \\
at $\lambda$=360 nm          &  \%    &   $\geq$35   &   $\geq$10 & $\geq$35   &  $\geq$25 &  $\geq$35       \\
at $\lambda$=420 nm          &  \%    &   $\geq$60   &    $\geq$55 & $\geq$60   &  $\geq$55 &  $\geq$60                          \\
at $\lambda$=620 nm          &  \%    &   $\geq$70   &   $\geq$65 &  $\geq$70   & $\geq$65 &  $\geq$70             \\
\hline
Inhomogeneity of Transverse   &  nm    & $\leq$5      &   $\leq$6 &  $\leq$5      &  $\leq$3 &   $\leq$3                                   \\
Transmission $\Delta \lambda$ at T=50\% &     &              &    & & &                                            \\
\hline
Induced radiation absorption & m$^{-1}$ & $\leq$1.1   &  $\leq$1.5 & $\leq$1.1   & $\leq$1.6 &   $\leq$1.1                   \\
coefficient $dk$ at $\lambda$=420 nm &       &     &        & &     &             \\
and RT, for integral dose $\geq$100 Gy &     &       &     & &    &                \\
\hline   
 Mean value of $dk$          & m$^{-1}$  & $\leq$0.75 &      & $\leq$0.75 &    &                            $\leq$0.75            \\
\hline
Tolerance in Length          & $\mu$m   & $\leq\pm$150 &   -100/+300 &  $\leq\pm$150 & $\leq\pm$100 &        $\leq\pm$50      \\
Tolerance in sides          & $\mu$m   & $\leq\pm$50 &   $\pm 0$ & $\leq\pm$50 &  $\leq\pm$50 &                 $\leq\pm$50                     \\
\hline
Surface polished, roughness Ra & $\mu$m & $\leq$0.02 &   & $\leq$0.02 & $\leq$0.02 &          \\
\hline   
Tolerance in Rectangularity (90$^o$) & degree & $\leq$0.1 &   & $\leq$0.1 & $\leq$0.12 & $\leq$0.01  \\
\hline
Purity specific. (raw material) &    &        &          & &  &       \\
Mo contamination                & ppm & $\leq$1 &  &  $\leq$1 & $\leq$10 &   $\leq$1             \\
La, Y, Nb, Lu contamination     & ppm & $\leq$40 &   & $\leq$40 &  $\leq$100 &  $\leq$40               \\
\hline
\end{tabular}\par}
\end{table*}

}

\section{Growth and production of crystals}
\label{sec-crystals}

The quality of scintillator material, e.g. crystals, depends strongly on the production process and associated quality assurance. In this section we review the benefits and limitations of production methods for PbWO$_4$ crystal growth and their implementation at the only two vendors with mass production capability of such materials worldwide.

\subsection{Crystal growth methods}
\label{growth-meth}

Crystal growth can roughly classified into three groups: solid-solid, 
liquid-solid and gas-solid processes, depending on which phase transition is 
involved in the crystal formation. The liquid-solid process is one of the 
oldest and widely used techniques. Crystal growth from melt is the most 
popular method. 
                             
The Bridgman technique~\cite{Bridge1925} is one of the oldest method used for growing crystals. The principle of the Bridgman technique is the directional solidification by translating a melt from the hot zone to the cold zone of the furnace. At first the polycrystalline material in the crucible needs to be melted completely in the hot zone and be brought into contact with a seed at the bottom of the crucible. This seed is a piece of single crystal and ensures a single-crystal growth along a certain crystallographic orientation.  

The crucible is then translated slowly into the cooler section of the furnace. 
The temperature at the bottom of the crucible falls below the solidification 
temperature and the crystal growth is initiated by the seed at the melt-seed 
interface. After the whole crucible is translated through the cold zone the 
entire melt converts to a solid single-crystalline ingot.

The Bridgman technique can be implemented in either a vertical or a horizontal system configuration~\cite{Bridge1925,Bridge-horiz1,Bridge-horiz2}. The concept of these two configurations is similar. The vertical Bridgman technique enables the growth of crystals in circular shape, unlike the D-shaped ingots grown by horizontal Bridgman technique. However, the crystals grown horizontally exhibit high crystalline quality and lower defect densities, since the crystal experiences lower stress due to the free surface on the top of the melt and is free to expand during the entire growth process.

The Czochralski process~\cite{Czoch1918,Czoch-alt} is a method of crystal growth used to obtain 
single crystals. It take a seed of future crystal and attach it to the stick, 
then slowly pulled up the stick (0.5-13 mm/h) by rotating it in the same time. 
The crucible may, or may not, be rotated in the opposite direction. 
The seed will grow into much bigger crystal of roughly cylindrical shape. 
The seed should be an oriented single crystal. 
The Czochralski process is more difficult, and is good for congruently melting 
materials (oxides, silicon among others). By precisely controlling the temperature gradients, rate of pulling and speed 
of rotation, it is possible to extract a large, single-crystal ignot from the 
melt. This process is normally performed in an inert atmosphere, such as argon,
and in an inert chamber, such as quartz. 
Large variety of semiconductors and crystals, including PbWO$_4$ can be grown 
by this method.
                            
The Czochralski method is one of the major melt-growth techniques. 
It is widely used for growing large-size single crystals for a wide range of 
commercial and technological applications. One of the main advantages of 
Czochralski method is the relatively high growth rate. 

\subsection{Brief description of PbWO$_4$ crystal history}
\label{crystal-hist}

Mass production of PbWO$_4$ was developed by CMS in order to produce the 
crystals required for use at LHC. During the CMS and early PANDA EMC 
construction, two manufacturers, Bogoroditsk Technical Chemical Plant (BTCP) in Russia and The Shanghai Institute of Ceramics of the Chinese Academy of Sciences (SICCAS) in China, using different crystal growth methods were available. Essentially all high quality crystals have been produced at BTCP using the Czochralski growing method, whereas SICCAS produces crystals using the Bridgman method. BTCP is now out of business, and the worldwide availability of high quality PbWO$_4$ production has changed dramatically. 

SICCAS produced 1825 crystals out of the about 70k crystals for the CMS electromagnetic calorimeter (EMCal), 1200 crystals for the JLab Hybrid EmCal, and a few hundred crystals for the PANDA EMCal project between 2011 and 2015. SICCAS has produced $\sim$670 crystals for the NPS project between 2014 and 2019. The characterization of these crystals is described in the following sections. 

The only other producer with mass production capability for PbWO$_4$ in the world is CRYTUR in the Czech Republic. CRYTUR started work on PbWO$_4$ at the end of 1995, considerably later than BTCP and SICCAS, and did not play a major role during the CMS EMCal construction. CRYTUR returned its focus on PbWO$_4$ production in the early 2010's through collaborations with PANDA and EIC. CRYTUR is using the Czochralski crystal growing method and has been using the pre-production crystal materials from BTCP as raw material. CRYTUR is expected to produce all $\sim$ 8000 crystals for the PANDA EMCal barrel approximately 700 crystals for the NPS. About 350 crystals for the NPS project have been delivered between 2018 and 2019. The characterization of these crystals is described in the following sections.

\section{Crystal Quality Assurance}
\label{sec-quality-assur}

Quality assurance and control of the scintillator material is important for high precision physics measurements and also an important part of the production process. Measurement of properties important for physics can provide feedback for optimizing material formulation and fabrication process. The acceptable limits for the NPS in comparison to those for EIC and other projects are listed in Table~\ref{tab:pwo-specification}.

\subsection{Samples}
\label{subsec-samples}

\begin{figure}[H]
\centering
{\includegraphics[width=3.5in]{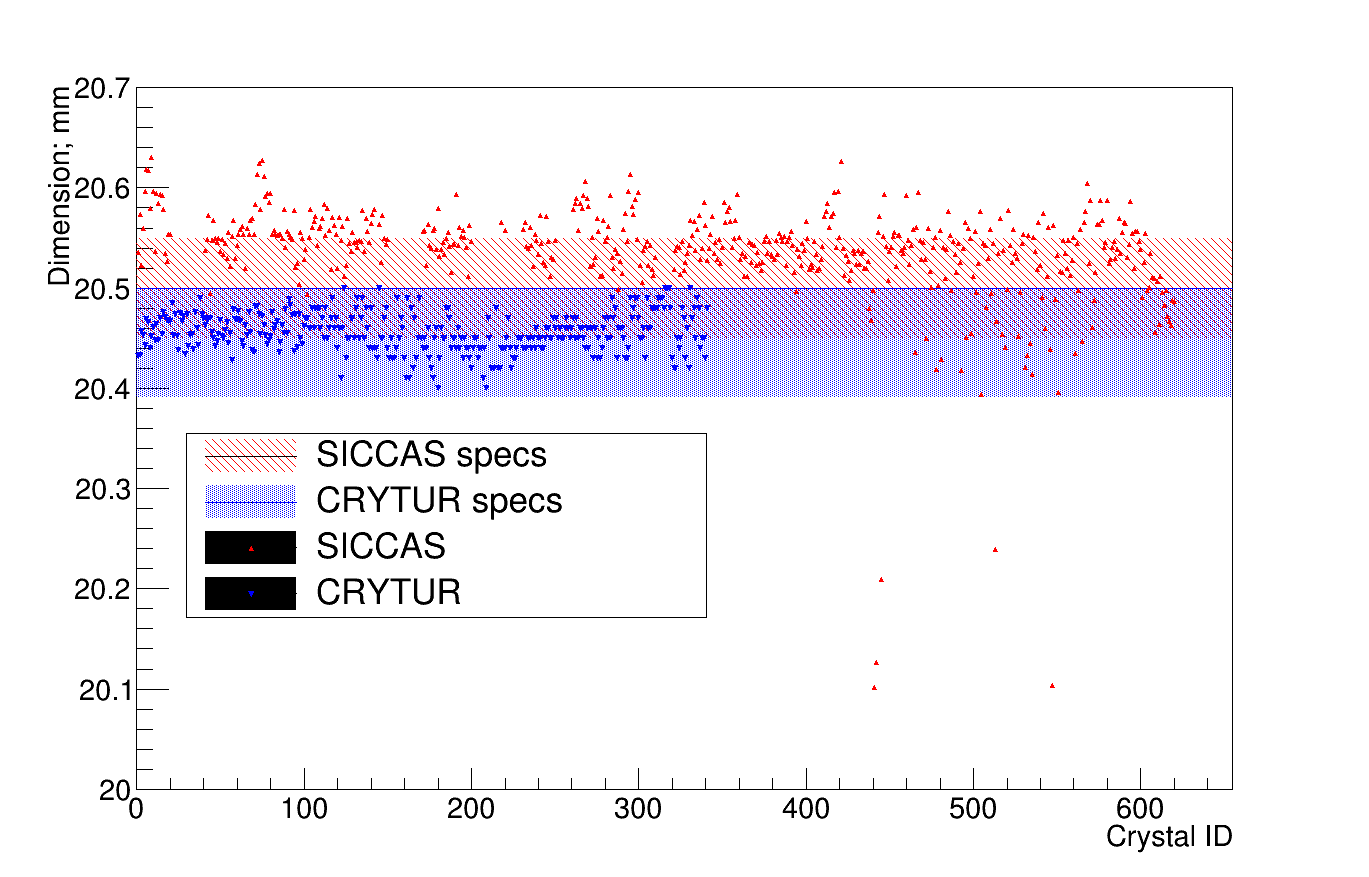}}
\caption{\label{fig:dim_all} (Color online) The measured dimensions of the crystals.}
\end{figure}

A total of 350 PbWO$_4$ samples from Crytur and 666 PbWO$_4$ samples from SICCAS were studied in this investigation. The samples had rectangular shape. Their nominal dimensions are 2.05 cm x 2.05 cm x 20 cm. The longitudinal and transverse dimensions of all samples were measured using a Mitutoyo Electric Digital Height Gage ($\sim1~\mu$m accuracy). Table~\ref{tab:crystal-dim} lists the average dimensions, year of production, crystal grower, and production technology for all samples, and Fig.~\ref{fig:dim_all} shows the measured dimensions for a subset of 529 SICCAS and 311 Crytur crystals.

\begin{table*}
\caption{\label{tab:crystal-dim} PbWO$_4$ crystal dimensions.}
{\centering  \begin{tabular}{|c|c|c|c|}
\hline
Vendor & Production Technology   & Year of Production & Average dimensions     \\
\hline
Crytur & Czochralski   & 2018 &  200.00 $\pm$ 0.01, 20.470 $\pm$ 0.019         \\
Crytur & Czochralski   & 2019 &  200.00 $\pm$ 0.01, 20.460 $\pm$ 0.015         \\
SICCAS & Bridgman   & 2014 & 200.0 $\pm$ 0.2, 20.0 $\pm$ 0.02                  \\
SICCAS & Bridgman   & 2015 & 200.5 $\pm$ 0.2, 20.1 $\pm$ 0.02                  \\
SICCAS & Bridgman   & 2017/18 & 200.0 $\pm$ 0.2, 20.550 $\pm$ 0.025            \\
SICCAS & Bridgman   & 2019 & 200.0 $\pm$ 0.2, 20.540 $\pm$ 0.027               \\
\hline
\end{tabular}\par}
\end{table*}

All crystals from Crytur were grown by the Czochralski method. Crystals Crytur-001 to Crytur-100 were produced in 2018, crystals Crytur-101 to Crytur-350 were produced in 2019. All samples from SICCAS were grown using the modified Bridgeman method. Crystals SIC-01-15 were produced in 2014, crystals SIC-16-45 in 2015, crystals SIC-046-506 in 2017/18, and crystals SIC-506 to SIC-666 in 2019. All samples from Crytur were transparent and clear without major voids and scattering centers visible to the eye. A few samples were found to be cloudy, which was traced back to the polishing equipment. One sample had a yellow film, which was found to be leftover polishing solution. Samples from SICCAS showed yellowish, brownish, and pink color. The yellow color may be caused by absorption bands in the blue region. Many of the SICCAS samples had macroscopic voids and scattering centers visible to the eye and highlighted under green laser light. Microscopic defects and voids not visible to the eye are discussed in section~\ref{subsec:Surface-property}. All surfaces of the samples were polished by the manufacturer and no further surface treatment, other than simple cleaning with alcohol, was carried out before the measurements. Samples were received without any irradition exposure. To test the impact of annealing for new crystals, SICCAS samples SIC-001 to SIC-045 and 50 samples of SIC-046 to SIC-506 were characterized before and after thermal annealing. 

\subsection{Optical transmission}

The longitudinal transmission was measured using a double-beam optical spectrometer with integrating sphere (Perkin-Elmer Lambda 950) in the range of wavelengths between 200 and 900 nm. The systematic uncertainty of transmittance was better than 0.3\%. The  reproducibility of these measurements is better than 0.5\%. 

Additional uncertainties in the transmittance measurement arise due to the birefrigent nature of PbWO$_4$ crystals and due to macroscopic defects, e.g. voids, inclusions, scattering centers. The uncertainty due to birefrigence was estimated to be less than 10\% for different azimuthal angle orientations of the crystal. For the main measurements the crystal was set up at a specific azimuthal angle, which gave the maximum longitudinal transmittance. The major contribution to uncertainty in many SICCAS samples was due to macrodefects. The effect was minimized by using an integrating sphere, which collected almost all light passing through the sample, and collimation of the light path to maximize the longitudinal transmittance. 

If one assumes that light impinges normally on the crystal surface and that the two end surfaces are parallel, one can determine the average light attenuation length using~\cite{Zhu96},
\begin{equation}
L_{attenuation}=\frac{l}{ln \frac{T(1-T_i)^2)}{\sqrt{4T^2_i + T^2(1-T^2_i)^2-2T_i^2}}}
\end{equation} 
where $l$ is the length of the crystal, $T$ is the measured transmittance, and $T_i$ is the real theoretical transmittance limited only at the end surfaces of the crystal. Taking into account multiple reflections,
\begin{equation}
T_i= \frac{1-R}{1+R}
\end{equation}
where $R=(n-n_{air})^2/(n+n_{air})^2$ with $n$ and $n_{air}$ the refractive indices of PbWO$_4$ and air, respectively. 

\begin{figure}[H]
\centering
{\includegraphics[width=3.0in]{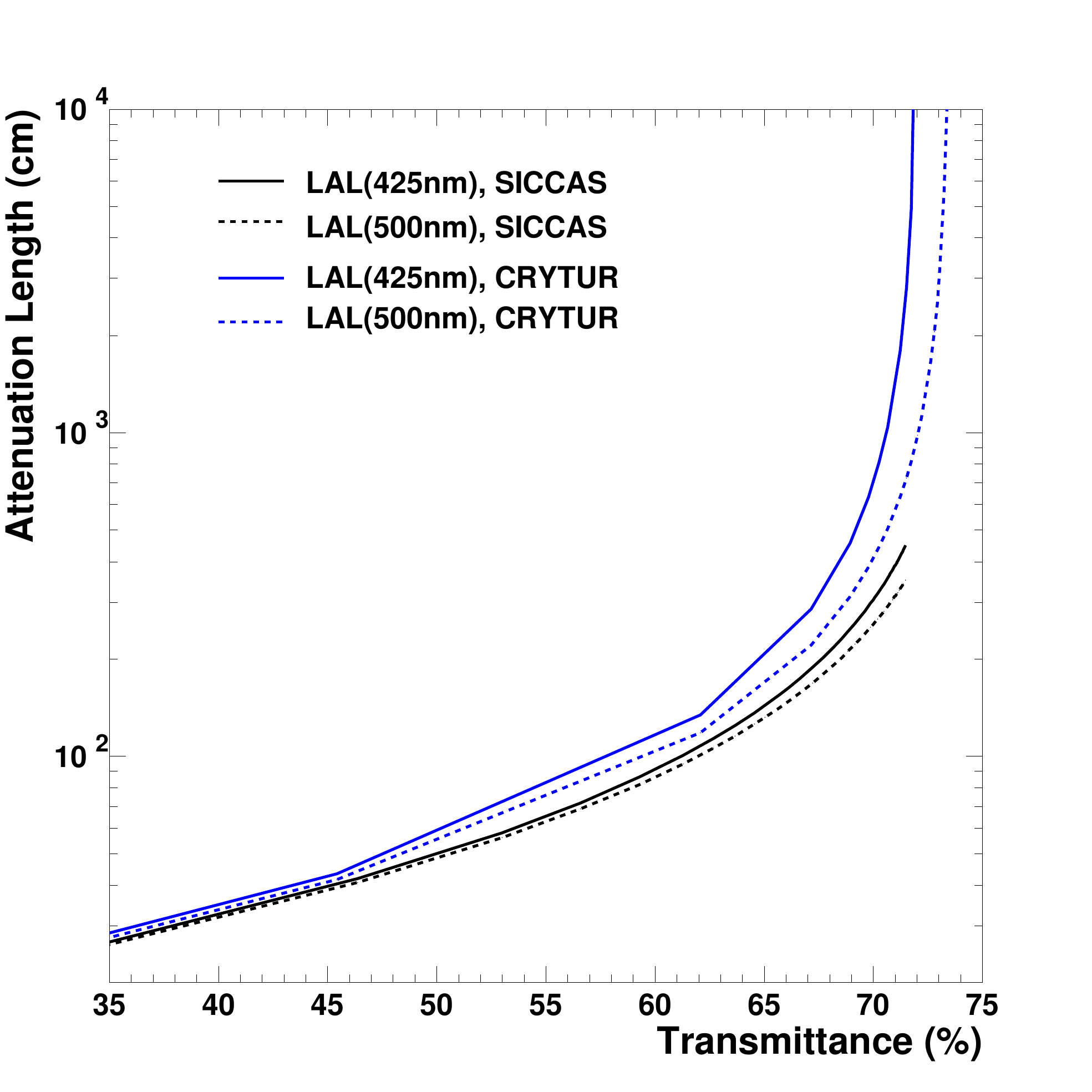} }
\caption{\label{fig:attenuation-length} (Color online) Attenuation length at 425nm (solid) and 500nm (dashed) for CRYTUR (blue) and SICCAS (black) crystals using the PbWO$_4$ extraordinary refractive index from Ref.~\cite{Baccaro}.}
\end{figure}

The light attentuation length of Crytur and SICCAS crystals at 425 and 500 nm calculated using the PbWO$_4$ extraordinary refractive index from Ref.~\cite{Baccaro} is shown in Fig.~\ref{fig:attenuation-length}. 

\begin{figure}
\centering
\subfigure[\label{fig:Transmittance-setup}] 
%caption for subfigure a
{\includegraphics[width=1.5in]{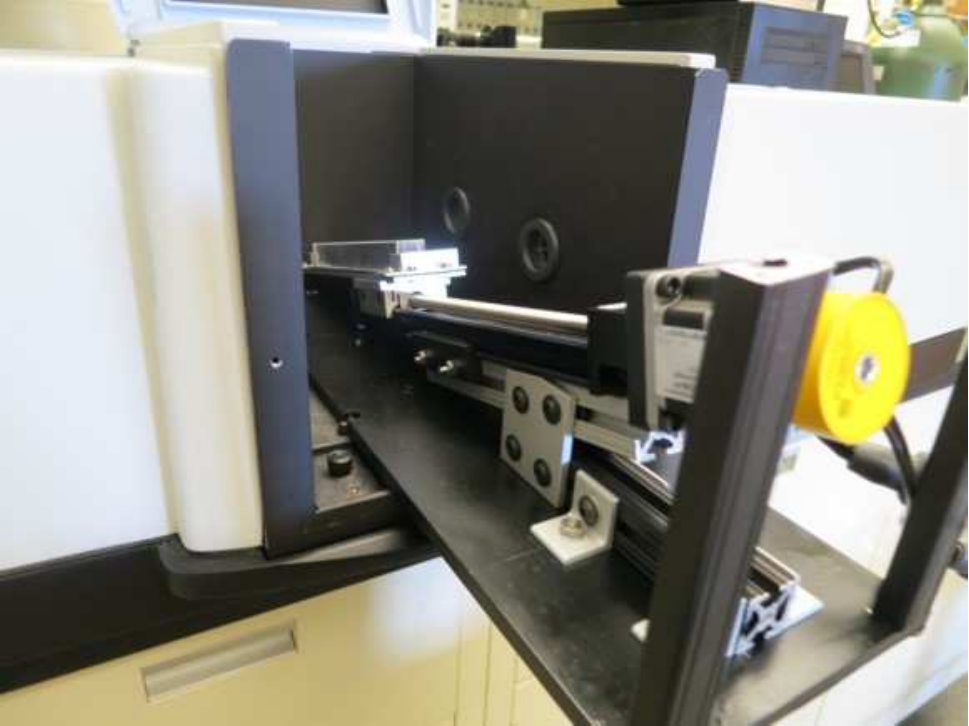} }
\hspace{0.1cm}
\subfigure[\label{fig:Transmittance-3D}]
% caption for subfigure b
{\includegraphics[width=1.5in]{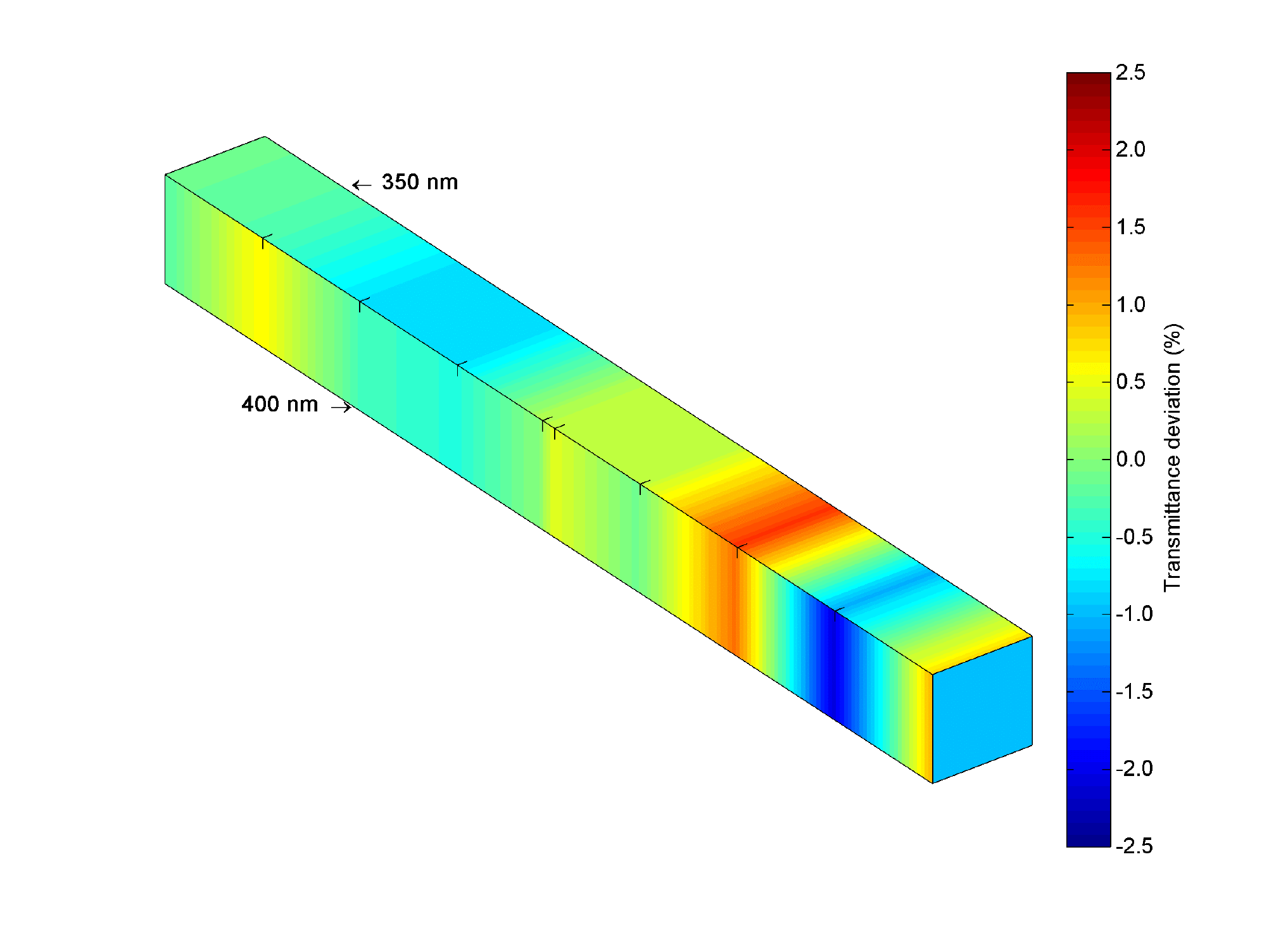} }
\caption{\label{fig:Transmittance_setup} (Color online) Left: Modification to spectrophotometer for transverse transmittance measurements. Right: 3D transmittance map of a crystal. The low transmittance regions are due to bubbles in the volume.}
\end{figure}

The homogeneity of the crystal is investigated based on the 
variation of the transverse optical transmission. A quality parameter that 
characterizes the band edge absorption of the crystal is defined as the 
maximum variation of the wavelength at a transmission value of T=50\% along the length of the crystal. In addition, the maximum \% deviation of the transverse transmission from the value measured at the center are used. 
Both, the transverse optical absorbance and the longitudinal transmission were measured as function of wavelength to characterize the crystal quality.

\subsection{Luminescence yield, temperature dependence and decay kinetics}

\begin{figure}
\centering
{\includegraphics[width=\columnwidth,height=3.8in]{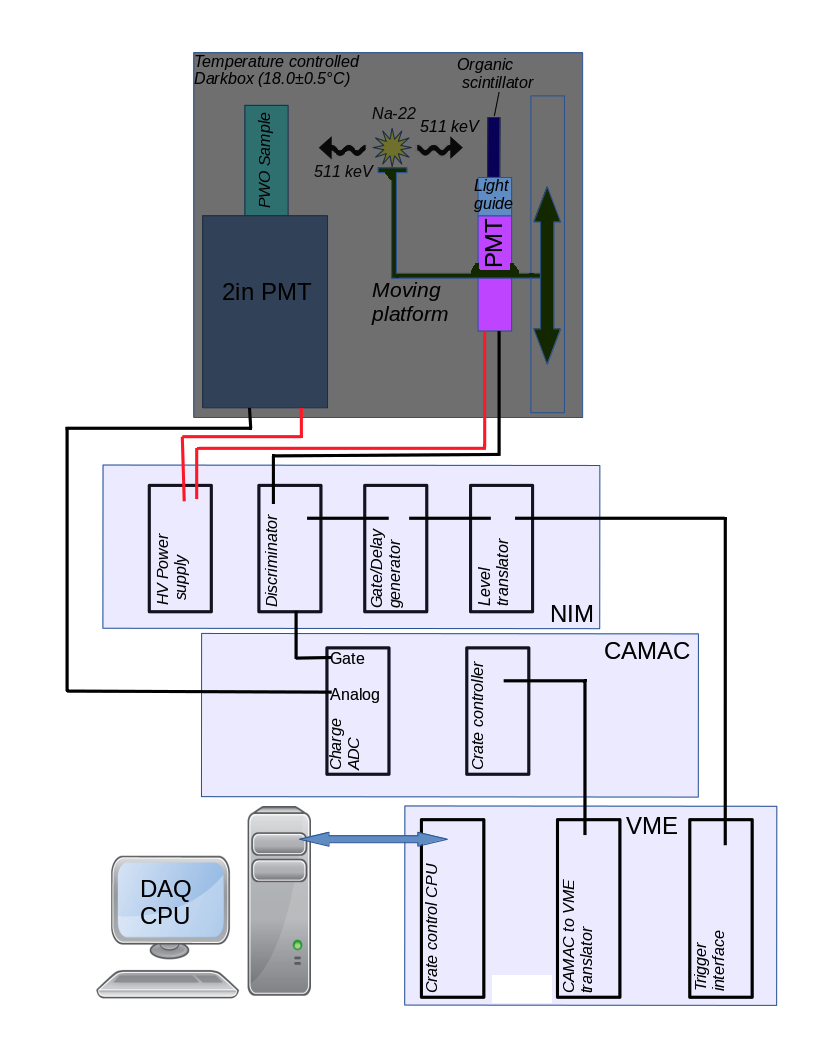}}
\caption{\label{fig:ly-setup} (Color online) Schematic of the light yield measurement setup inside a temperature-controlled darkbox.}
\end{figure}

The scintillation light yield at 18 degrees Celsius was determined at CUA using a $^{22}$Na source emitting back-to-back photons of 0.511 keV from $e^-e^+$ annihilation (see Fig.~\ref{fig:ly-setup}). One of the end faces of the crystal was optically coupled to the entrance window of a 2-inch photomultiplier tube (Photonis XP2282, quantum efficiency $\sim$27\% at 400nm) using Bicron BC-630 optical grease. All other surfaces of the crystal were wrapped in three layers of Teflon film and two layers of black electrical tape. The anode signals were directly digitized using a charge sensitive 11 bit integrating type analog-to-digital converter (ADC LeCroy 2249W) with integration gates between 100 ns and 1000 ns, to investigate the contribution of slow components. The effective integration gate for the main measurements was 150 ns. The photoelectron number corresponding to the $\gamma$ source peak was determined from the peak ADC channel obtained with a Gaussian fit. To calibrate the signal amplitude above the pedestal in units of photoelectrons a separate measurement was made to determine the response to a single photoelectron. 

At fixed light intensity the number of detected photoelectrons depends only on the PMT quantum efficiency, $QE \propto N_{pe}$. Neglecting contributions from electronic noise and other possible fluctuations the $N_{pe}$ can be estimated as inverse square of the normalized width of the detected photoelectron distribution,
\begin{equation}
 N_{pe}=1/\sigma^2_{norm}, 
\end{equation}
where $\sigma_{norm}=\sigma/N_{ADC}$, with $\sigma$ the width of the amplitude distribution determined from a Gaussian fit and $N_{ADC}$ is the pedestal subtracted signal amplitude in ADC channels.
 
The setup is operated inside a temperature-controlled dark box, which provides for temperature accuracy and stability on the order of better than 1$^o$C. The dependence of the light yield on the temperature was measured to be 2.4\%/$^o$C. This is consistent with previous measurements published in Ref.~\cite{Semenov}. 

%The light response uniformity was measured by moving the photon source along the length of the crystal. 

To determine the setup dependence of the light yields, subsets of crystals were characterized at Orsay, as well as the facilities at Giessen U. and Caltech. The Orsay facility uses a $^{137}$Cs source. Crystals are wrapped in four layers of teflon, 1 layer of aluminum foil, and a black heat shrinking tube. The open end is coupled to the entrance window of a 2-inch photomultiplier tube (Photonis XP5300B) with QE peak around 29\%. The anode signals were digitized using a Desktop Digitizer 5730 with effective integration gate 150 ns and full range up to 1000 ns. At the Giessen facility crystals are excited with 662 keV photons from a $^{137}$Cs source. Crystals are wrapped in eight layers of teflon, 1 layer of aluminum foil, and black heat shrinking tube. The open end is coupled to a 2-inch PMT (Hamamatsu R2059-01) with typical quantum efficiency 20\% at 420nm. The PMT signal above a suitable threshold was integrated in time gates of 100 ns to 1000 ns and digitized wih a Charge-to-Digital-Converter (CAMAC, Le Croy 2249W). The Caltech facility uses the same sources as Orsay and Giessen. The light was detected with a Hamamatsu R2059 PMT with quartz window. Crystals were wrapped in one layer of Tyvek paper or 5 layers of teflon. Measurements were typically made at 23$^\circ$C, while measurements at CUA, Orsay, and Giessen are made at 18$^\circ$C.

A major difference that affects the absolute number of photoelectrons measured with each setup is the quantum efficiency of the PMTs as discussed in Ref.~\cite{Mao}. The gamma-ray excited luminescence of PWO shows a broad and complex emission band ranging from 370 to 500 nm. The shape of the emission spectrum can be correlated with the specific conditions of the crystal synthesis, e.g. the tungsten concentration in the melt~\cite{Korzhik}. We thus focus here on the correlations of the measurements between setups rather than absolute values. 

The scintillation decay was evaluated by measuring the light yield as a function of the integration gate. This allows for analyzing the relative contribution of slow components. If such slow components contribute significantly an increase in the relative light yield beyond 1000 ns should be clearly visible. In general, the light yield increases by a factor of about three due to cooling to -25 $^o$C independent of the integration time window. 

\subsection{Gamma ray irradiation}

\begin{figure}
\centering
{\includegraphics[width=3.5in]{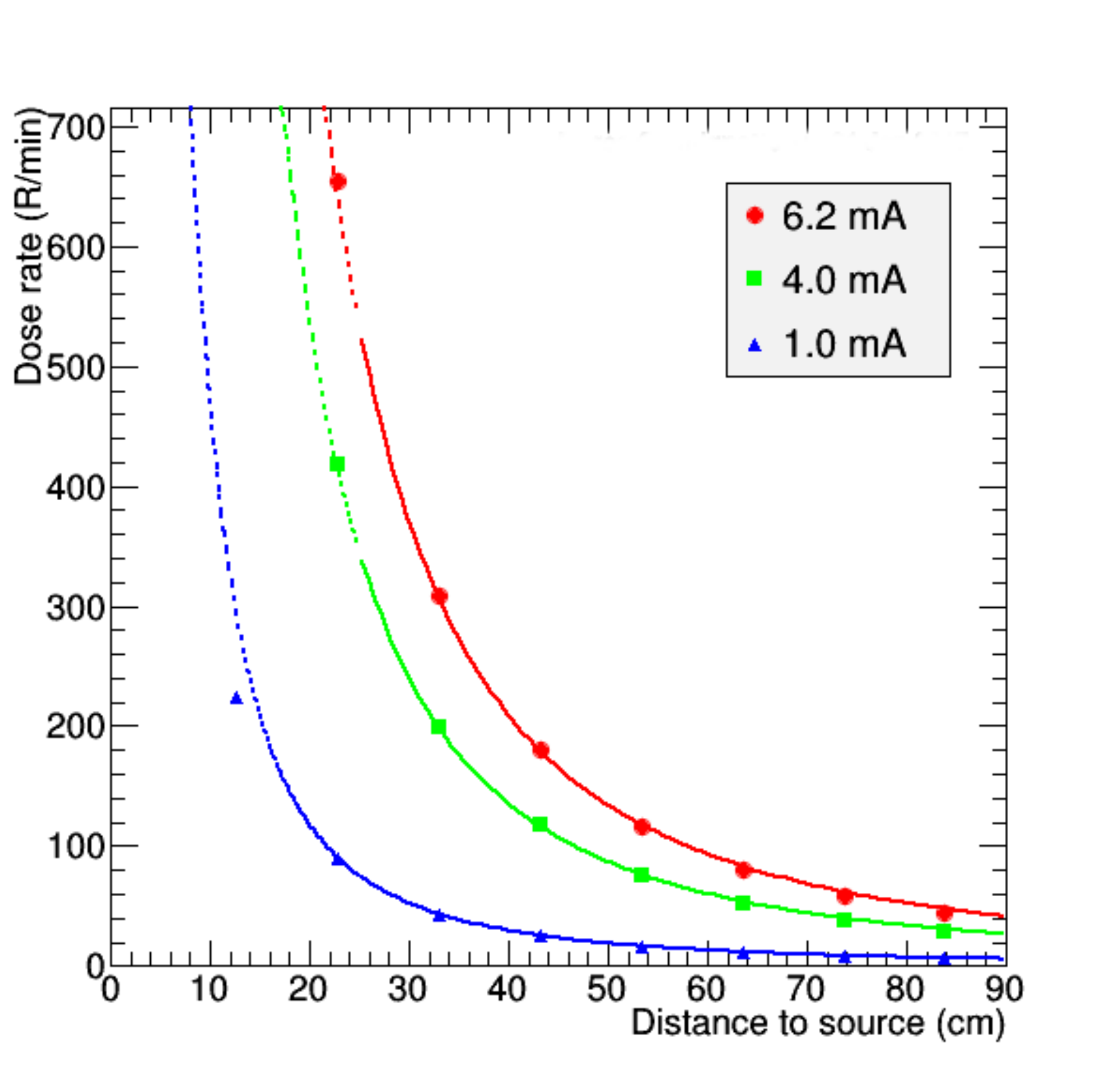}}
\caption{\label{fig:xray-dose-rate} (Color online) The Faxitron CP160 Xray dose rate as function of distance from the source.}
\end{figure}

\begin{figure}
\centering
\subfigure[\label{fig:Radiation_Xray}] 
%caption for subfigure a
{\includegraphics[width=1.6in]{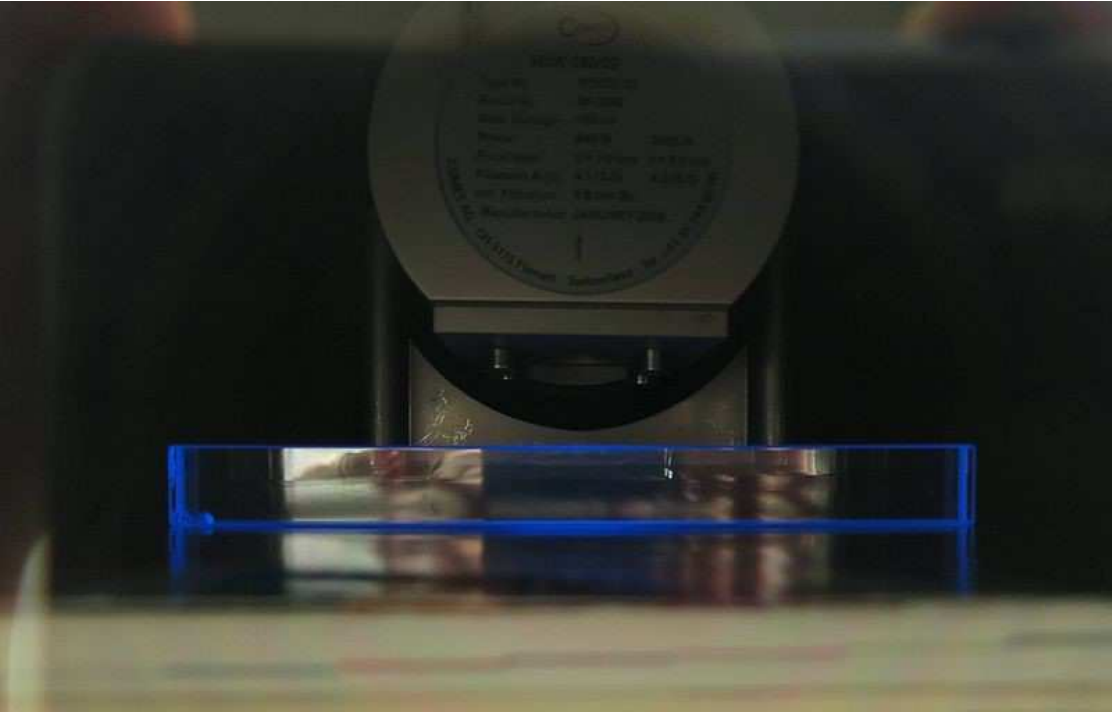} }
\hspace{0.1cm}
\subfigure[\label{fig:Radiation_Xray_damage}]
% caption for subfigure b
{\includegraphics[width=1.4in]{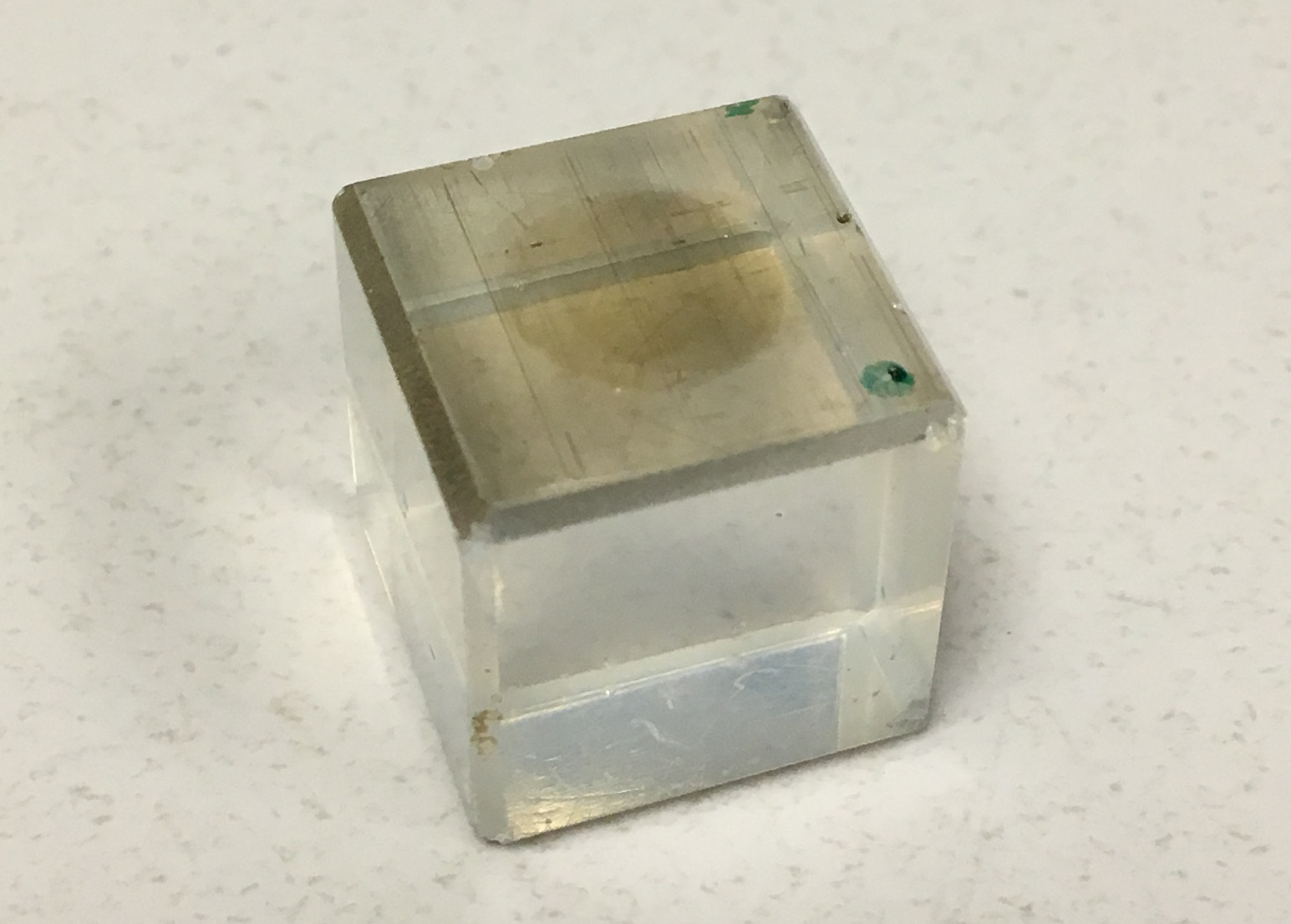} }
\caption{\label{fig:Radiation_Xray_examples} (Color online) Left: Crystal irradiated by Xrays; Right: Example of radiation damage induced by Xrays and integrated dose of 1000 Gy.}
\end{figure}

The irradiation tests were carried out at two different facilities to provide a cross check between measurements. The first was carried out at CUA using the cabinet X-ray (Faxitron CP160). The optical transmittance was determined before and after irradiation with integral doses of 30-100 Gy imposed within an irradiation period of 10 minutes. The crystals were kept light tight during and after irradiation until the transmission measurement commenced to minimize the effect of optical bleaching. The measurement was performed no later than 30 minutes after the end of the irradiation procedure at room temperature. The dose rates (see Fig.~\ref{fig:xray-dose-rate}) were determined using a RaySafe ThinX dosimeter and data provided by the manufacturer. The dose rate at a current of 6.2mA was parameterized as Dose rate (R/min) = (-8537 + 55720*Current)/Distance to source, where the distance to the source varies between 22.9cm and 83.8cm. The parameterization can be converted to Gy using the conversion factor 0.00877. The dose rate uncertainty is estimated to be 2\% for currents 6.2~mA. The Xray photon radiation damage manifests at the surface of the crystal. An example is shown in Fig.~\ref{fig:Radiation_Xray_examples}.
 
The second irradiation facility was the Laboratoire de Chimie Physique in Orsay. This facility features a panoramic irradiation complex based on 2 $^{60}$Co sources with a total activity of 2000 Ci. Crystals were irradiated with integrated doses ranging from 500 Gy to 1000 Gy at about 18 Gy/min. The dose rate was accurately measured using Fricke dosimetry, which consists of measuring the absorption of light produced by the increased concentration of ferric ions by ionizing radiation in a solution containing a small concentration of ammonium iron sulfate. The linear absorption with time at a given position determines the exact radiation dose received by the crystal when placed at the same position as the solution. PbWO$_4$ crystals were irradiated to 30 Gy at 1 Gy/min. 
%Preliminary results indicate an average value of the radiation-induced absorption coefficient of 0.7 m$^{-1}$ with values ranging from 0.4 to 1.1 m$^{-1}$.

\begin{figure}
\centering
{\includegraphics[width=3.0in]{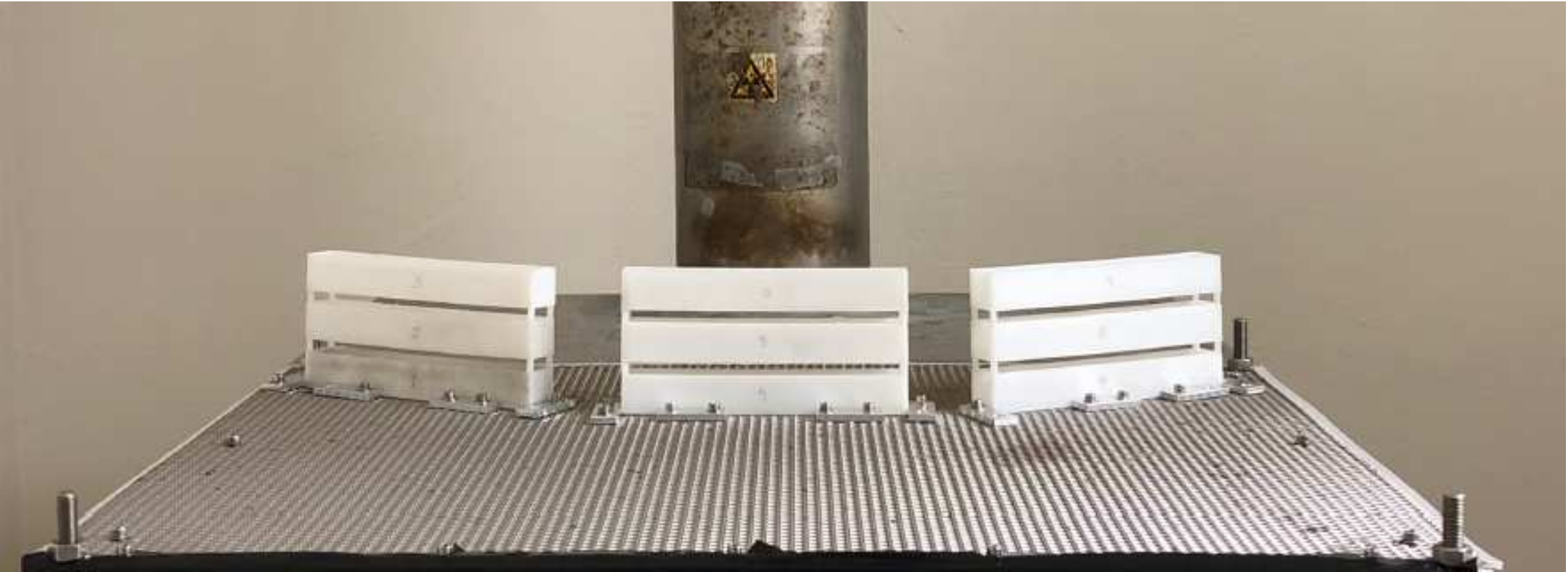} }
\caption{\label{fig:Radiation_Co60_setup} (Color online) Irradiation setup with a high activity $^{60}$Co source. Crystals are placed in containers where the radiation dose was previously measured using a Fricke solution. }
\end{figure}

The $^{60}$Co source allowed for irradiating multiple crystals at the same time. To estimate the dose and dose rate in the crystals, a Fricke solution positioned at the same distance (60 cm from the source) and of the same shape and volume as the crystals was irradiated. 

\begin{figure}[H]
\centering
{\includegraphics[width=\columnwidth,height=2.5in]{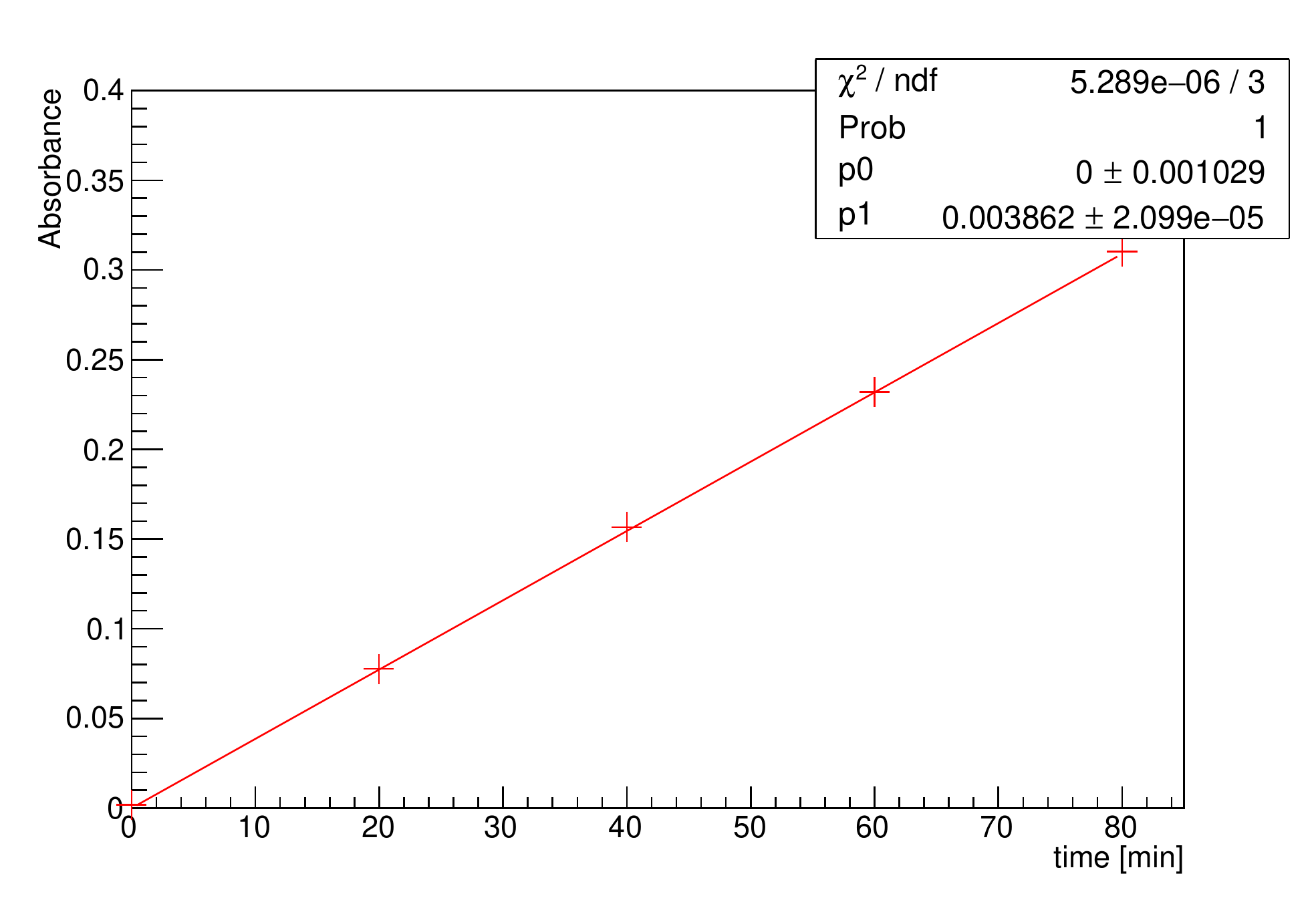}}
\caption{\label{fig:radiation-fricke-absorbance} (Color online) The measured absorbance vs. irradiation time in the Fricke solution.}
\end{figure}

Fricke dosimetry is well studied. It changes light absorption linearly under radiation at a given wavelength up to about 200 Gy. The mechanism is the oxidation of ferrous ions (Fe$^{2+}$) to ferric ions (Fe$^{3+}$). Ferric ions aborb light and this absorption increases as the dose increases. To quantify the dose rate, we measured the light absorption for different irradiation times at the absorption peak of 304 nm at a distance of 60cm from the source. The result is shown in Fig.~\ref{fig:radiation-fricke-absorbance}. The solution's absorbance can be calculated using
\begin{equation*}
A  =  log \frac{I}{I_0} = \epsilon \times l \times C
   =  \epsilon \times l \times G \times \rho \times D(t)
\end{equation*}
where $I$ is the measured light intensity through the material, $\epsilon$ is the molar extinction coefficient (2160 + 15 (T-25) at 304 nm), $l$ is the optical path, $C$ is the number of moles transformed by the irradiation, $G$ is the radiolytic yield for $Fe^{3+}$ formation (1.62 $\times$ 10$^{-7}$ mol/J), $\rho$ is the mass density of the solution, and $D(t)$ is the radiation dose. The dose rate in Gray per minute is then given by,
\begin{equation*}
D(t)= \frac{\Delta A (cm^{-1})}{\epsilon (L mol^{-1}) \times G (mol J^{-1}) \times \rho (kg L^{-1}) \Delta t (min)}
\end{equation*}
The resulting average dose rate is 1.07 Gy/min with a standard deviation of 0.12 Gy/min.

The impact of radiation effects can be quantified in terms of the change in the absorption coefficient, $k$, which is determined from the longitudinal transmittance spectra before and after irradiation using
\begin{equation}
dk = \frac{ln(T_0/T_{rad})}{d}
\end{equation}
where $T_0$ and $T_{rad}$ are the measured transmittance before and after irradiation and $d$ is the total crystal length. The change in $k$ is shown over the entire spectrum of wavelengths in units of m$^{-1}$.

To quantify any setup dependent effects we carried out additional irradiation studies at Caltech and Giessen U. Caltech features a 4000 Ci $^{60}$Co source. Samples were irradiated at 2, 8, 30, 7000 rad/hour. The irradiation facility at the Giessen U Strahlenzentrum has a set of five $^{60}$Co sources. The homogeneity of the sources is on the level of 3.6 Gy/min. Samples are irradiated with an integral dose of 30Gy imposed within an irradiation period of 15 minutes. Crystals are kept ight tight during and after irradiation until transmission is started 30 min after the end of the irradiation. 

\subsection{Electron beam irradiation}

\begin{figure}
\centering
{\includegraphics[width=3.5in]{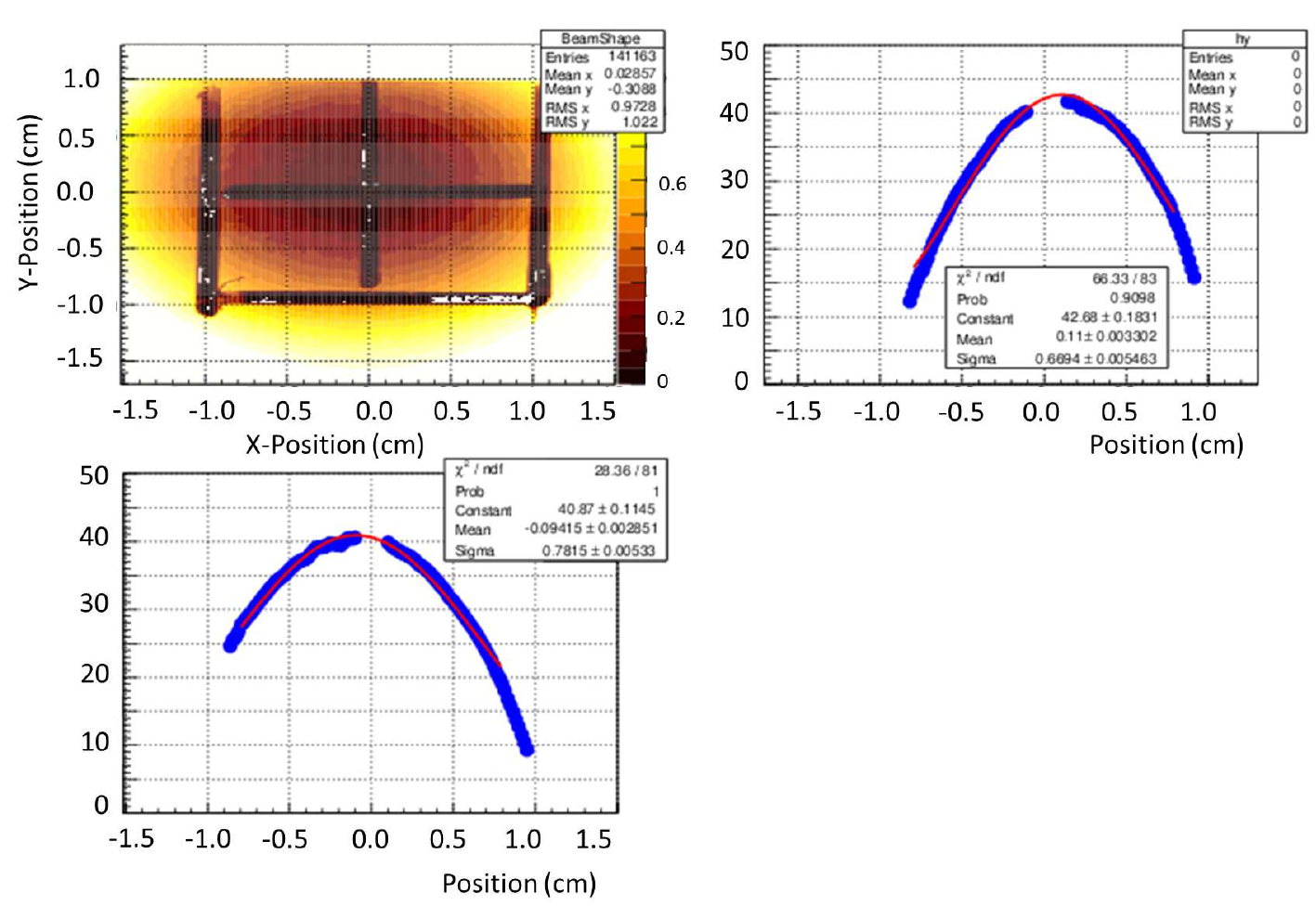} }
\caption{\label{fig:idaho-beam} (Color online) The glass plate exposed at the beginning of test at the Idaho Accelerator Facility (top left). Y (top right) and X (bottom left) profile of the beam at front plate located at 33 cm from the beam exit. Scanning and fitting give $\sigma_x$ $\sim$ 0.8 cm and $\sigma_y$ $\sim$ 0.7 cm). }
\end{figure}

The electron beam test was carried out at the Idaho Accelerator Facility, which features a 20 MeV electron beam with 100 Hz repetition rate and peak current $I_{peak}$=111 mA (11.1 nC per pulse and 100 ns pulse width). The beam is roughly 1 mm in diameter and exits through (1/1000) inch thick Ti window, a $x/X_0 = 7.1 \times 10^{-4}$ radiation length. Beam position and profile were measured using a glass plate. Scanning the plates and fitting the intensity distribution provides a quantitative (though approximate) measurement of the position and size of the beam at the location of the plate. The front plate was placed at the position of the PbWO$_4$ crystal front faces during irradiation that is 10.75 cm from the beam exit window. The rear plate was located at 33 cm from the beam exit, and shows the beam profile expansion. This provides a relatively homogeneous irradiation and heat load on the crystals. The beam profile is shown in Fig. \ref{fig:idaho-beam}. 

A PbWO$_4$ crystal at the above mentioned beam parameters has received a dose of 216 krad/min. Since such radiation dose rate is much higher ($\sim$13 Mrad/h) than the dose rates expected during the actual experiments, our tests were carried out at lower dose rates at a reduced  accelerator repetition rate, keeping the beam current per pulse and  pulse width unchanged. The measured relative difference of the crystal transmittance before and after irradiation is illustrated in Fig. \ref{fig:trans-degrad-432kr}. 
All transmittance measurements at the Idaho facility were carried out using an OCEAN OPTICS USB4000 device instead of a permanent spectrometer setup. The reproducibility of measurements with this setup ranges from 5\% to 15\%.

\section{Results of Crystal Characterization}
\label{sec-characterization-results}

\subsection{Transmittance and light attenuation length}
\label{subsec:optic-property}

\begin{figure}
\centering
\subfigure[\label{fig:Transmittance-Crytur}] 
%caption for subfigure a
{\includegraphics[width=3.0in]{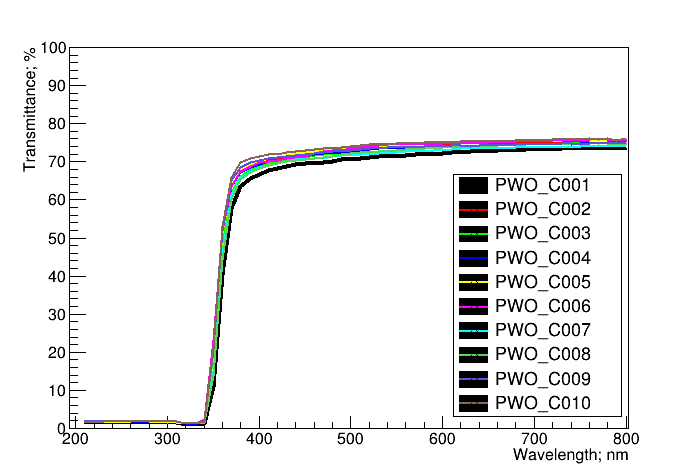} }
\subfigure[\label{fig:Transmittance-SICCAS}]
% caption for subfigure b
{\includegraphics[width=3.0in]{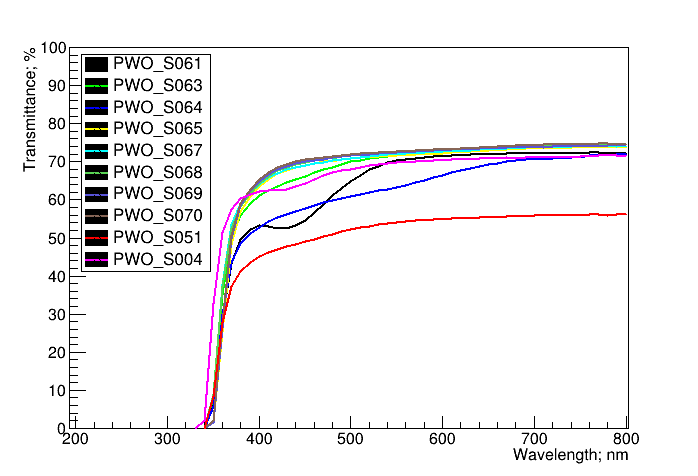} }
\caption{\label{fig:Transmittance-spectrum} (Color online) Representative longitudinal ransmittance spectra for Crytur crystals produced in 2018-19 (top) and SICCAS crystals produced in 2017 (bottom).}
\end{figure}

The longitudinal transmittance is shown in Fig. ~\ref{fig:Transmittance-spectrum}. Changes in the transmittance due to irradiation are discussed in section~\ref{sec-results-rad-damage}. 

The transmittance at 800 nm was $\ge$ 70\% for all Crytur and many SICCAS samples, and thus close to the theoretical limit. This implies a very long light attenuation length at this wavelength. No significant absorption was observed at wavelengths $>$ 550nm. For SICCAS samples with yellow, pink, or brown color significant absorption was observed below 550nm. The origin of the absorption is not understood. There are also considerable differences in transmittance spectra in the wavelength region between 350 and 550nm. Some SICCAS samples have a knee below 400nm, others show none. None of the Crytur samples show a knee. Samples with macro defects have very high transmittance at 360nm. The knee in the longitudinal transmittance can be correlated with radiation resistance. As discussed in section~\ref{sec-results-rad-damage}, samples irradiated with EM radiation and poor resistance will exhibit the knee below 400nm as well. 

\begin{figure}
\centering
{\includegraphics[width=3.5in]{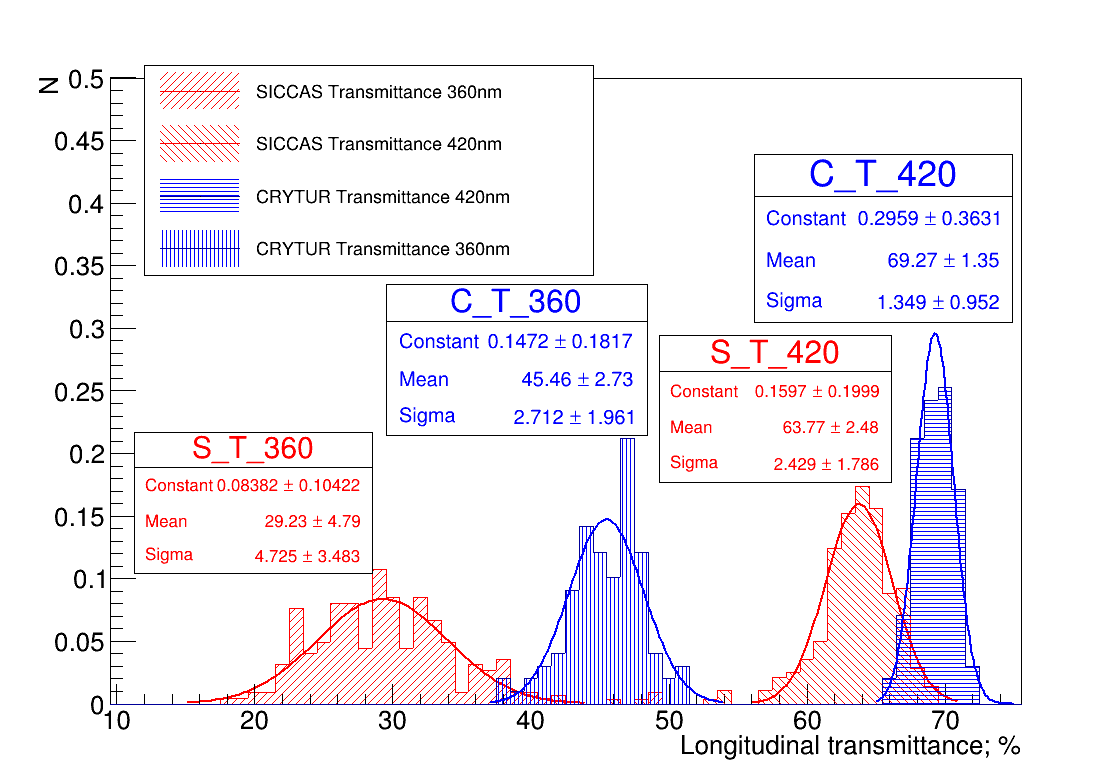}}
\caption{\label{fig:transmittance-longitudinal} (Color online) Longitudinal transmittance of Crytur and SICCAS crystals produced 2017-2019.}
\end{figure}

Fig.~\ref{fig:transmittance-longitudinal} illustrates the uniformity of the longitudinal transmittance for 150 Crytur and 150 SICCAS samples. CRYTUR crystals have an average transmittance of 69.3 $\pm$1.4 \% at 420nm and 45.5 $\pm$ 2.7 \% at 360nm. SICCAS crystals have an average transmittance of 64.0 $\pm$2.4 \% at 420nm and 29.2 $\pm$ 5.1 \% at 360nm. The broader distributions of the SICCAS crystals can be correlated with visual observation of mechanical defects, e.g. significant scattering centers in the bulk, as discussed in section~\ref{subsec-samples}. 

Compared to 23cm long crystals produced by SICCAS for CMS, the average performance of both Crytur and SICCAS crystals produced since 2014 is significantly improved. As published in Ref.~\cite{IEEE51}, the average longitudinal transmittance of CMS crystals is 21.3\%, 65.6\%, and 71.7\% at 360nm, 440nm, and 600 nm, respectively.

\begin{figure}
\centering
\subfigure[\label{fig:Transmittance-1D-good}] 
%caption for subfigure a
{\includegraphics[width=3.0in]{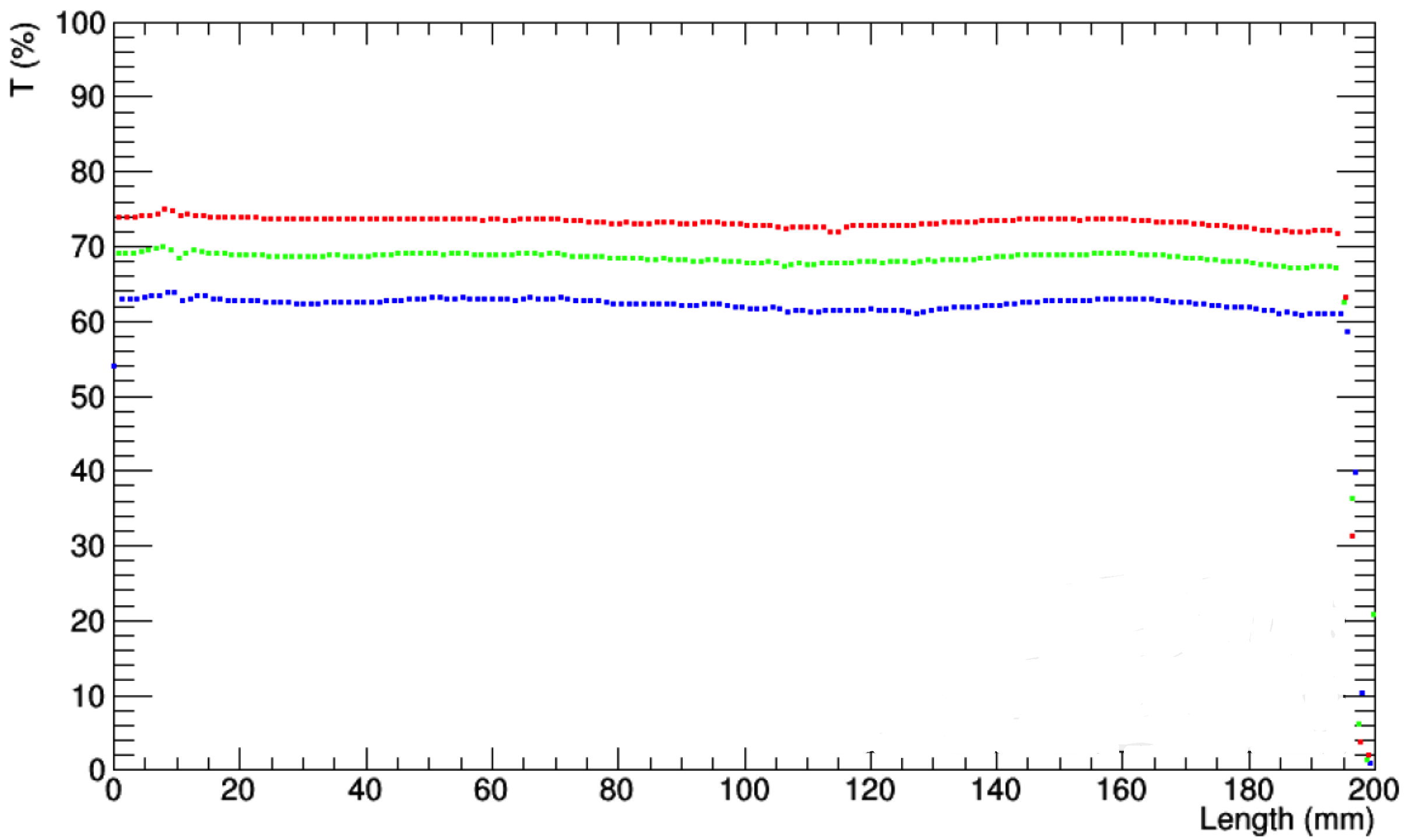} }
\hspace{0.1cm}
\subfigure[\label{fig:Transmittance-1D-bad}]
% caption for subfigure b
{\includegraphics[width=3.0in]{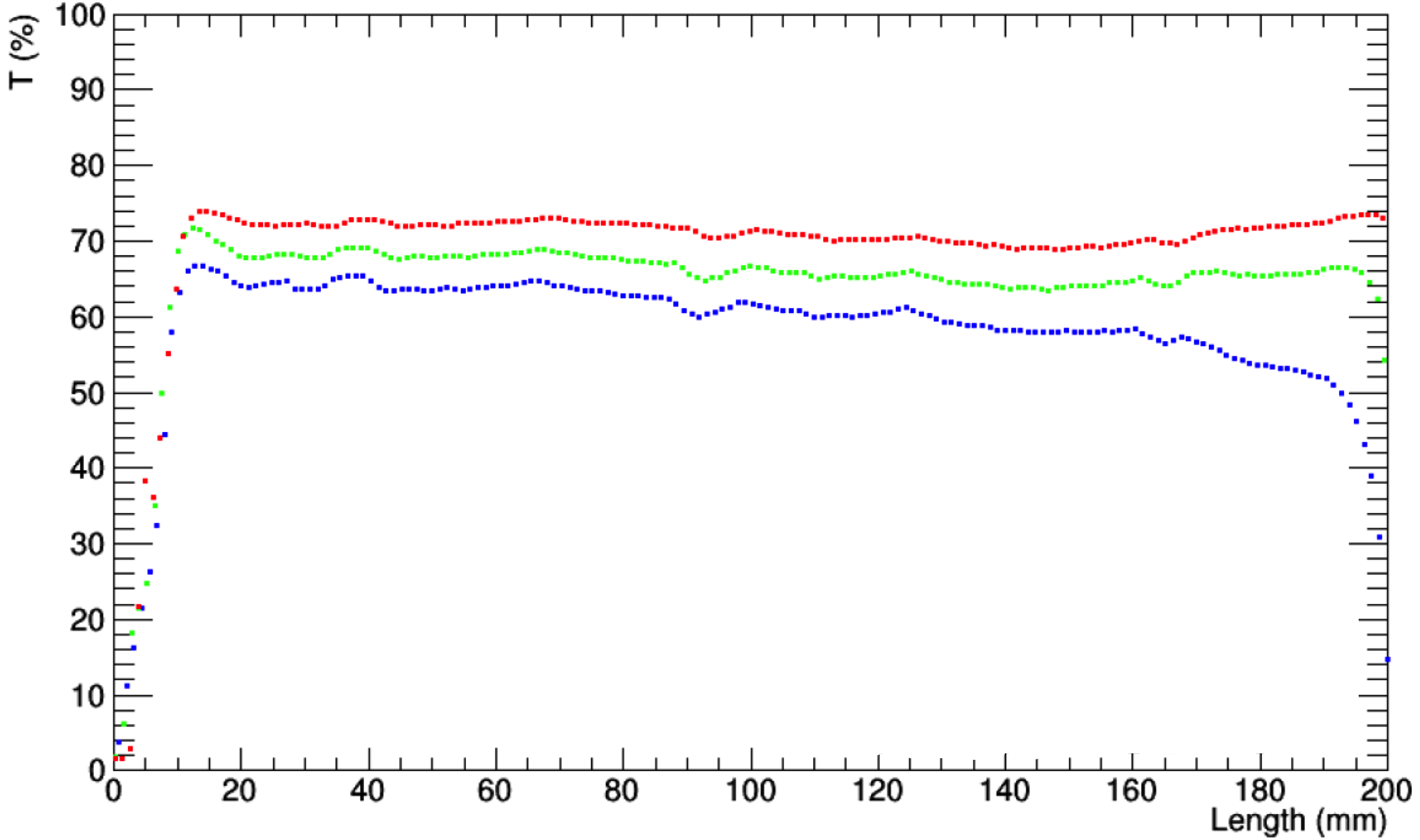} }
\caption{\label{fig:Transmittance_1D} (Color online) Transmittance transverse along the crystal for a (top) uniform and (bottom) nonuniform sample.}
\end{figure}

The transmittance in the transverse direction (2 cm thickness) was measured at several distances ranging between 5 and 195 mm from the face of 
the crystal. The results for one SICCAS crystal passing and one not passing specification are shown in Fig. \ref{fig:Transmittance_1D}. 

\subsection{Light Yield}
\label{subsec:light-yield}

\begin{figure}
\centering
{\includegraphics[width=3.5in]{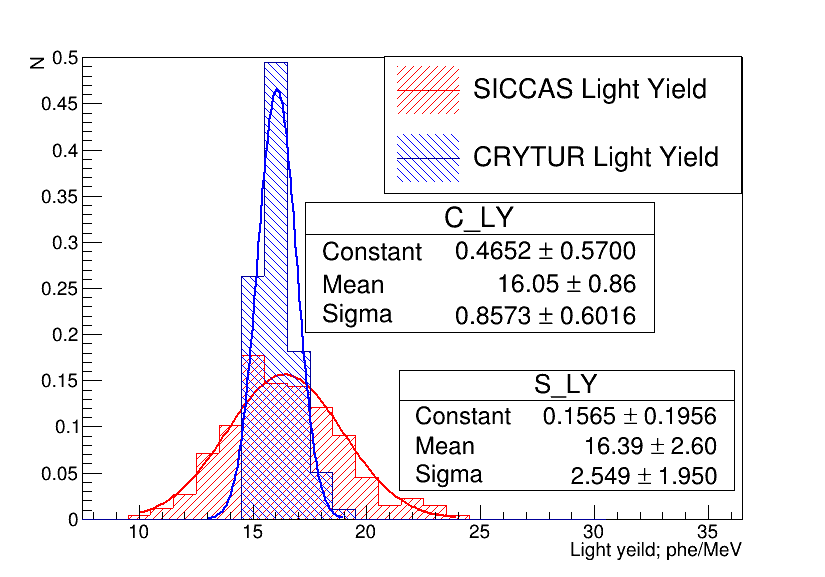}}
\caption{\label{fig:light-yield-all} (Color online) The measured light yield of the crystals.}
\end{figure}

The light yield of Crytur and SICCAS samples is shown in Fig.~\ref{fig:light-yield-all}. CRYTUR crystals have an average light yield of 16.1 with a variance of 0.9 photoelectrons/MeV, which is within the uncertainty of the measurement. SICCAS crystals have an average light yield of 16.4 with a variance of 2.6 photoelectrons/MeV. This large variation can be traced back to mechanical and chemical differences in crystals. 

\begin{figure}
\centering
\subfigure[\label{fig:LY-CUA-IPNO}] 
%caption for subfigure a
{\includegraphics[width=2.8in]{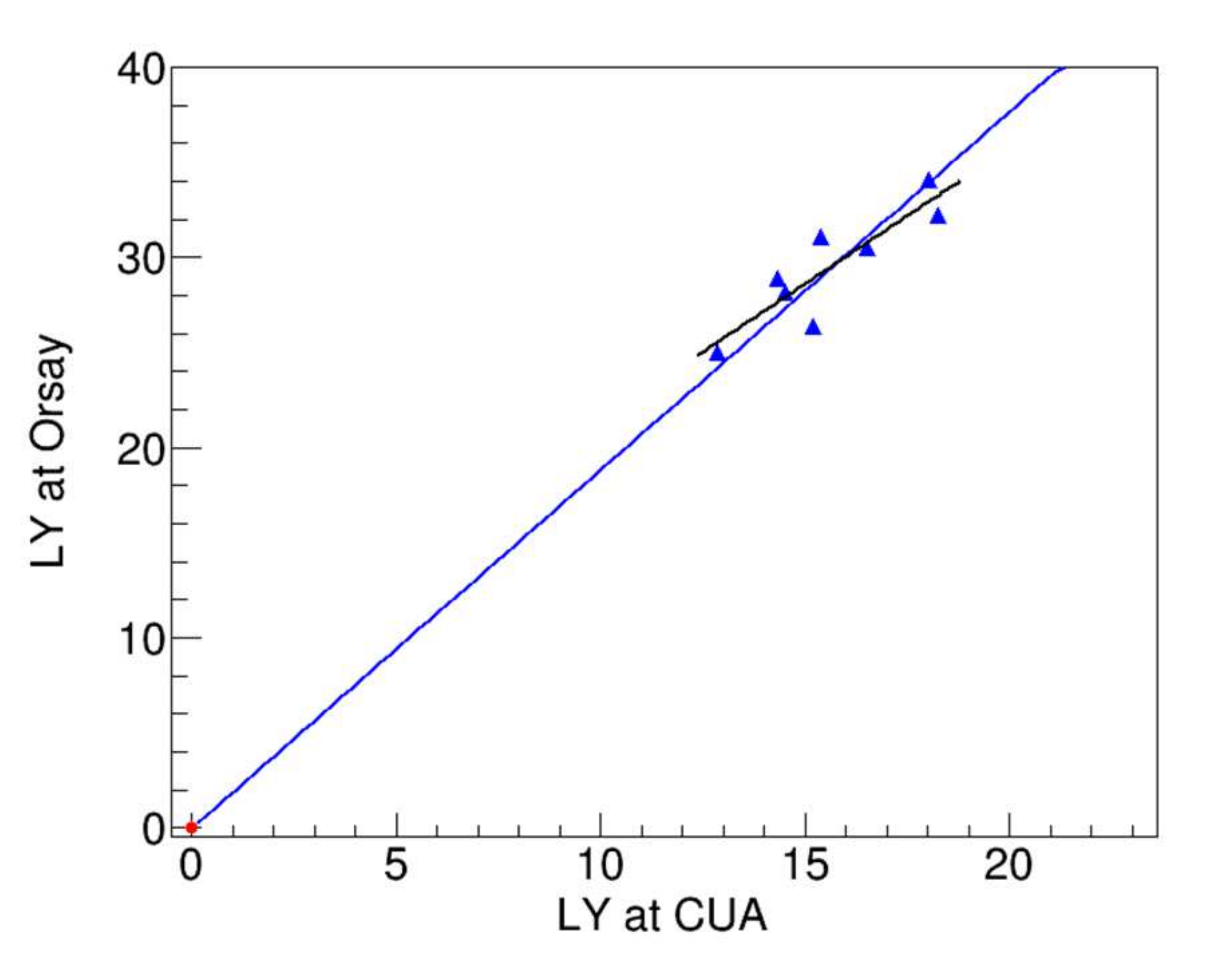} }
\hspace{0.1cm}
\subfigure[\label{fig:LY-CUA-Giessen}]
% caption for subfigure b
{\includegraphics[width=2.8in]{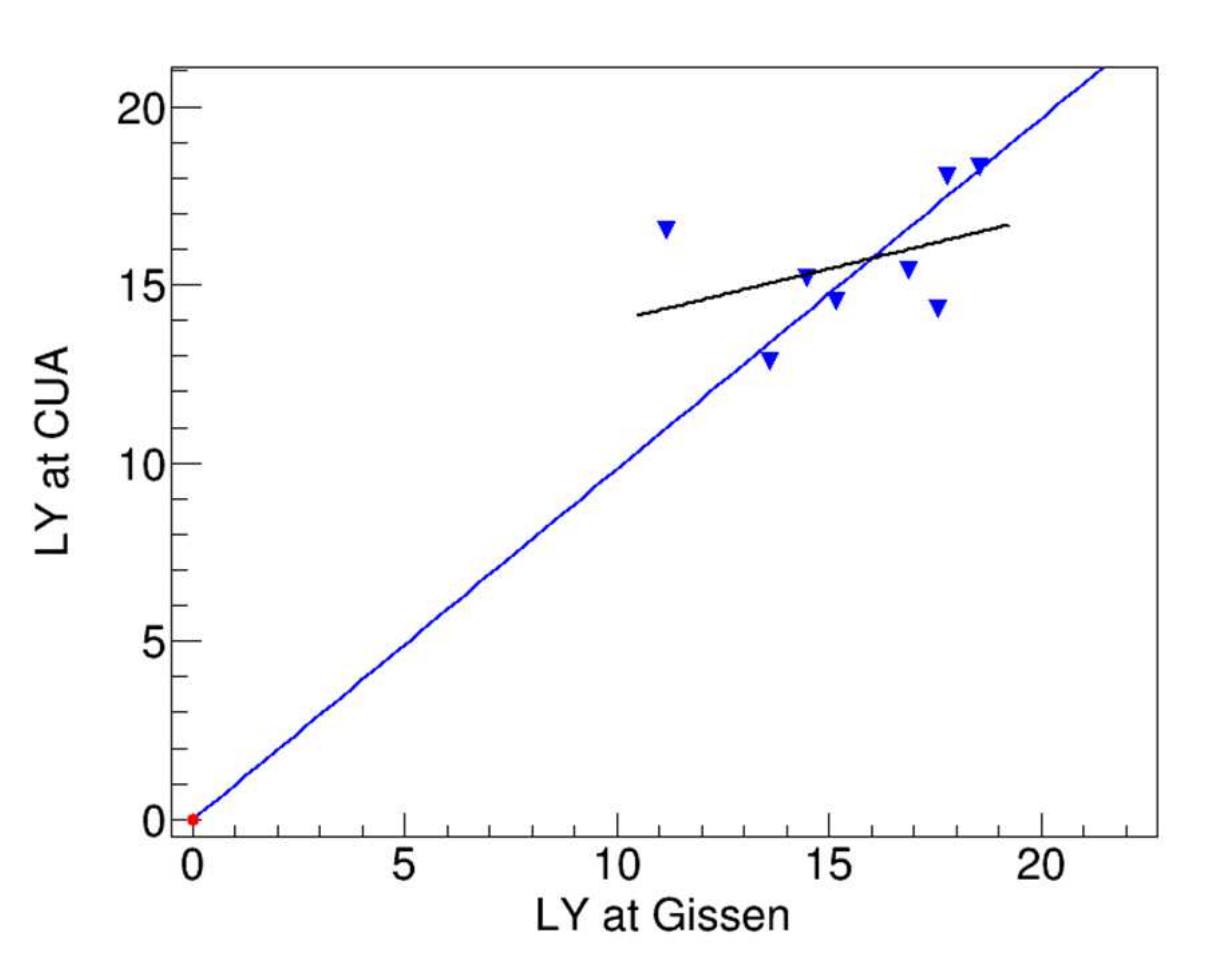} }
\hspace{0.1cm}
\subfigure[\label{fig:LY-IPN-Giessen}]
% caption for subfigure b
{\includegraphics[width=2.8in]{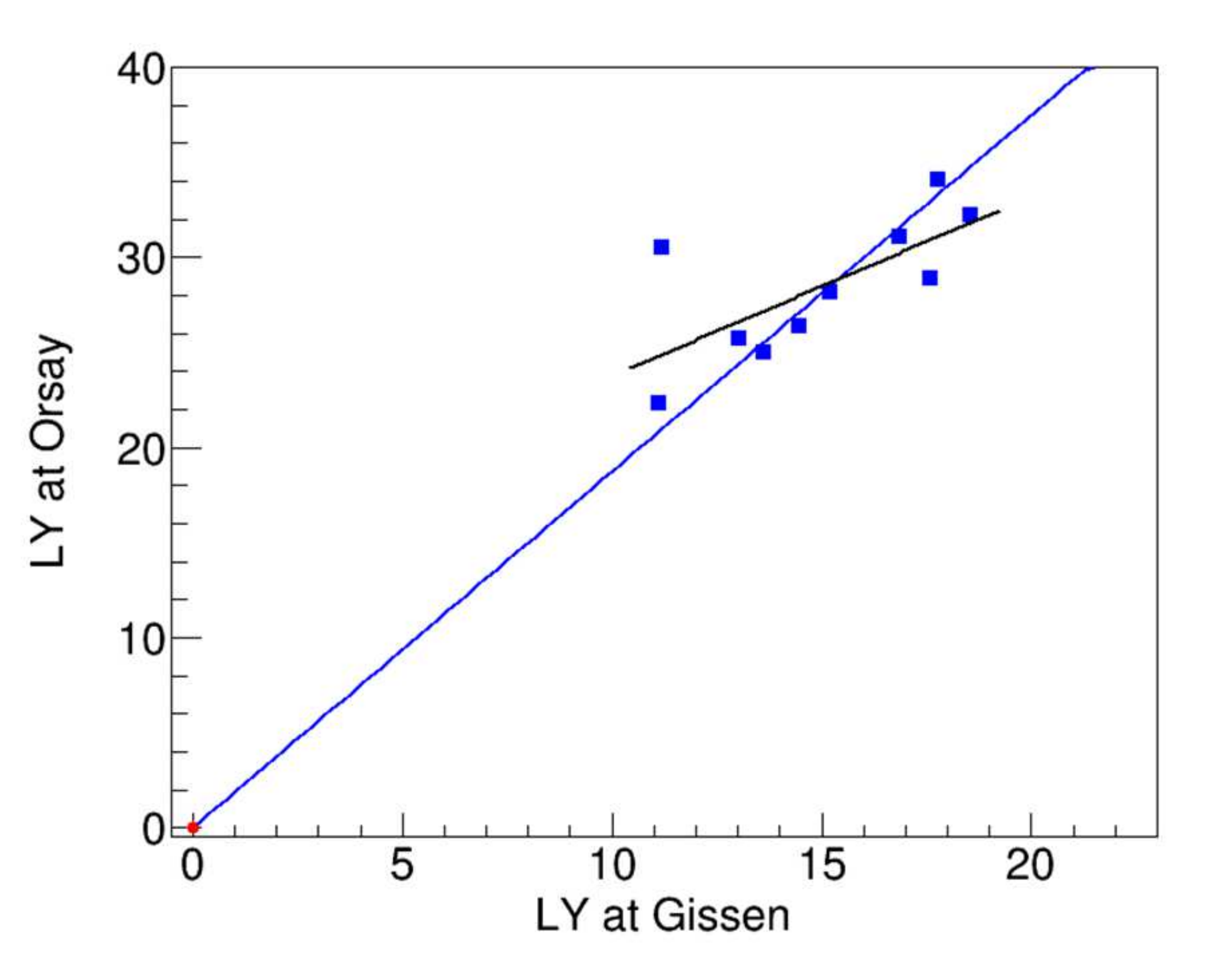} }
\caption{\label{fig:LightYield-Comparison} (Color online) Correlations between light yield measurements performed at CUA, Orsay, and Giessen. See text for details of each setup.}
\end{figure}

Measurement correlations between CUA, Orsay, and Giessen U. are shown in Fig.~\ref{fig:LightYield-Comparison}. The light yields of four crystals measured at Caltech and CUA agreed within one photoelectron. The absolute numerical values in photoelectrons to the vendor were given based on photoelectron numbers from the CUA setup.

Measurements done at Caltech also allowed for a direct comparison of crystals produced by SICCAS for CMS and since 2014 for the NPS project. All measurements were made at room temperature and with a 200ns gate. The average light output of 22x22x230 mm$^3$ PWO samples from CMS is 10.1 photoelectrons/MeV. In comparison, the 20x20x200 mm$^3$ PWO samples produced for NPS have an average light yield of 14.1 photoelectrons/MeV.

The light yield as a function of integration time was fitted to the parameterization
\begin{equation}
Light Output = A_0 + A_1*(1-e^{-t/\tau})
\end{equation}
where $A_0, A_1$ and $\tau$ are fit parameters. The fits show that over the time interval from 0 to 1000ns the decay times can be parameterized with a fast component, $\tau$ of 20 $\pm$ 1 ns. 

The scintillation decay kinetics is determined as the fraction of the total light output and the light yield integrated in a short time window of 100 ns. The measured values are on average 95\% for Crytur and 99\% for SICCAS crystals. The light yields for 100ns time windows are very similar and the fractional values are larger than 84\% and 96\% for CMS PWO crystals\cite{IEEE51}. 

\begin{figure}
\centering
{\includegraphics[width=\columnwidth,height=2.5in]{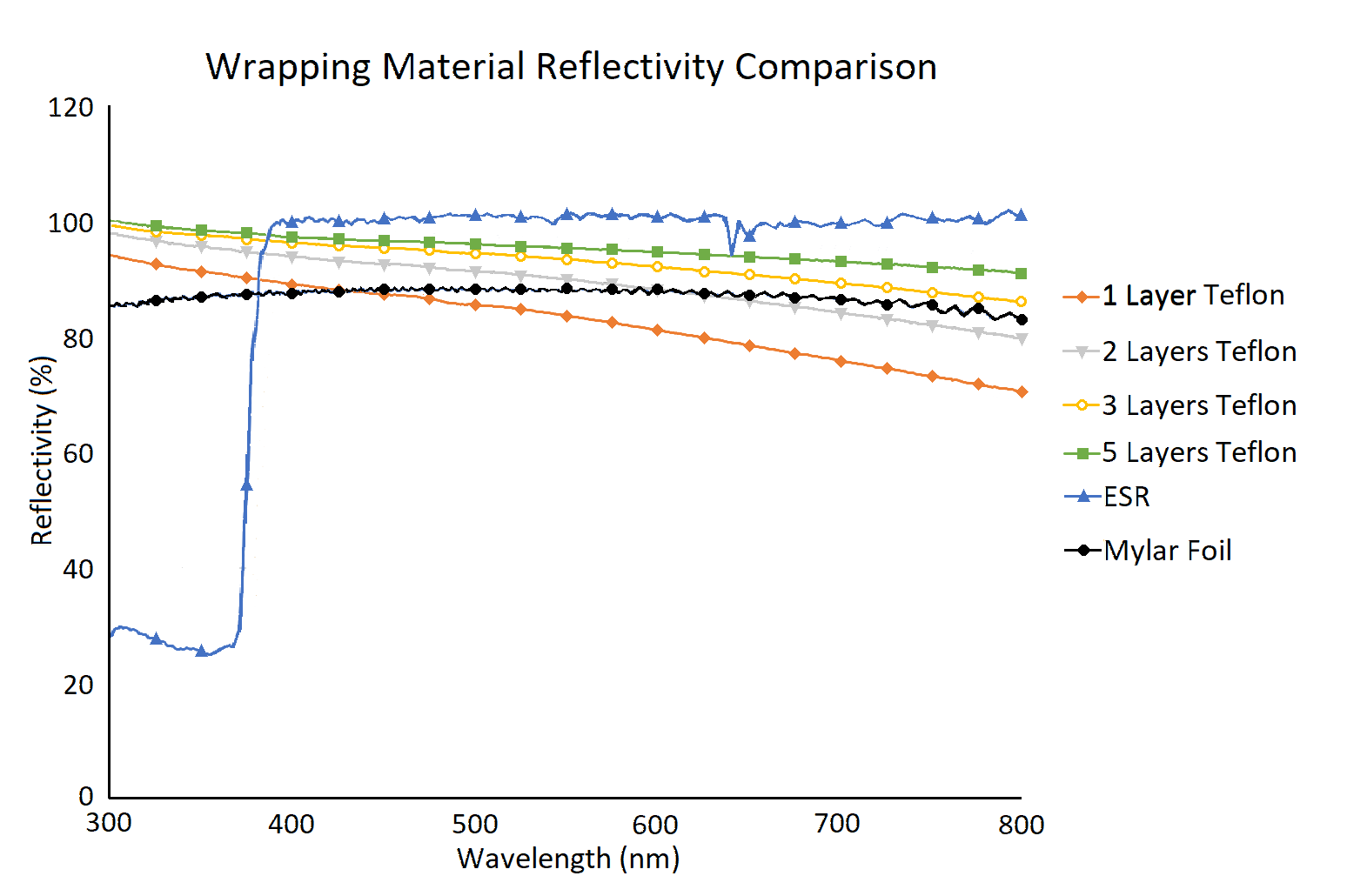}}
\caption{\label{fig:wrapper-reflectivity} (Color online) Reflectivity of mylar (black solid circles), teflon (1 (diamonds), 2 (upside down triangle), 3 (open circles), and 5 (squares) layers), and ESR (blue triangles).}
\end{figure}

%REFLECTOR
The performance of PbWO$_4$ crystal based calorimeter is highly dependent on the light-collection efficiency from the scintillator to the PMT. We have studied the effect of different reflectors and number of layers of reflectors on the light yield on PWO crystals. Fig.~\ref{fig:wrapper-reflectivity} shows the reflectivity of mylar, teflon, and Enhanced Specular Reflector (ESR) reflectors as measured with a spectrophotometer. 

Teflon tape is easily available and was our default choice for light yield tests. It is slightly transparent and therefore additional layers increase the reflectivity as shown in Fig.~\ref{fig:wrapper-reflectivity}. There is a clear positive trend from one to three layers, where the light yield increases significantly as the number of layers increases. The measured light yield follows the same trend as the reflectivity results. Three to four layers of teflon tape is thus the optimum amount. 

When used as a wrapping material, diffusive reflectors like teflon are more effective for light collection at 420 nm than specular reflectors. For example, mylar Foil produced lower light yields than 3 layers of Teflon Tape. On the other hand, Enhanced Specular Reflector produces the same light yield as three layers of teflon. The diffusive Gore reflector material has the highest reflectivity at 420 nm and also produced the highest light yield compared to both, three layers of teflon and ESR. Taking into account the mechanical properties of the reflector material and the constraints on total reflector thickness imposed by the detector design, the NPS uses one layer of 65$\mu$m ESR (VM2000). Tests were carried out to check for light cross talk between crystals and found no significant contamination.

It is interesting to note that the location of the reflector on the crystal has different importance for the total light collection. This was studied by comparing the light yield when the entire crystal was wrapped in 3 layers of Teflon Tape to those when only the bottom half (close to the PMT), the top half, small end face, or both end-and-top half were covered with reflector. The greatest impact on the light yield came from the reflector wrapped around the top half of the crystal resulting in a significant reduction of more than 8 photoelectrons in light yield when not present.

\section{Results on radiation damage}
\label{sec-results-rad-damage}

Possible effects of radiation damage in a scintillating crystal include radiation induced absorption, i.e. color center formation, effect on the scintillation mechanism, and radiation induced phosphorescence. Color center formation would affect the light attenuation length, and so the light output measured with the photodetector. Damage to the scintillation mechanism could affect the light output. Radiation induced phosphorescence could cause additional noise in the readout instrumentation. 

\subsection{Light Attenuation} 

\begin{figure}
\centering
{\includegraphics[width=3.0in]{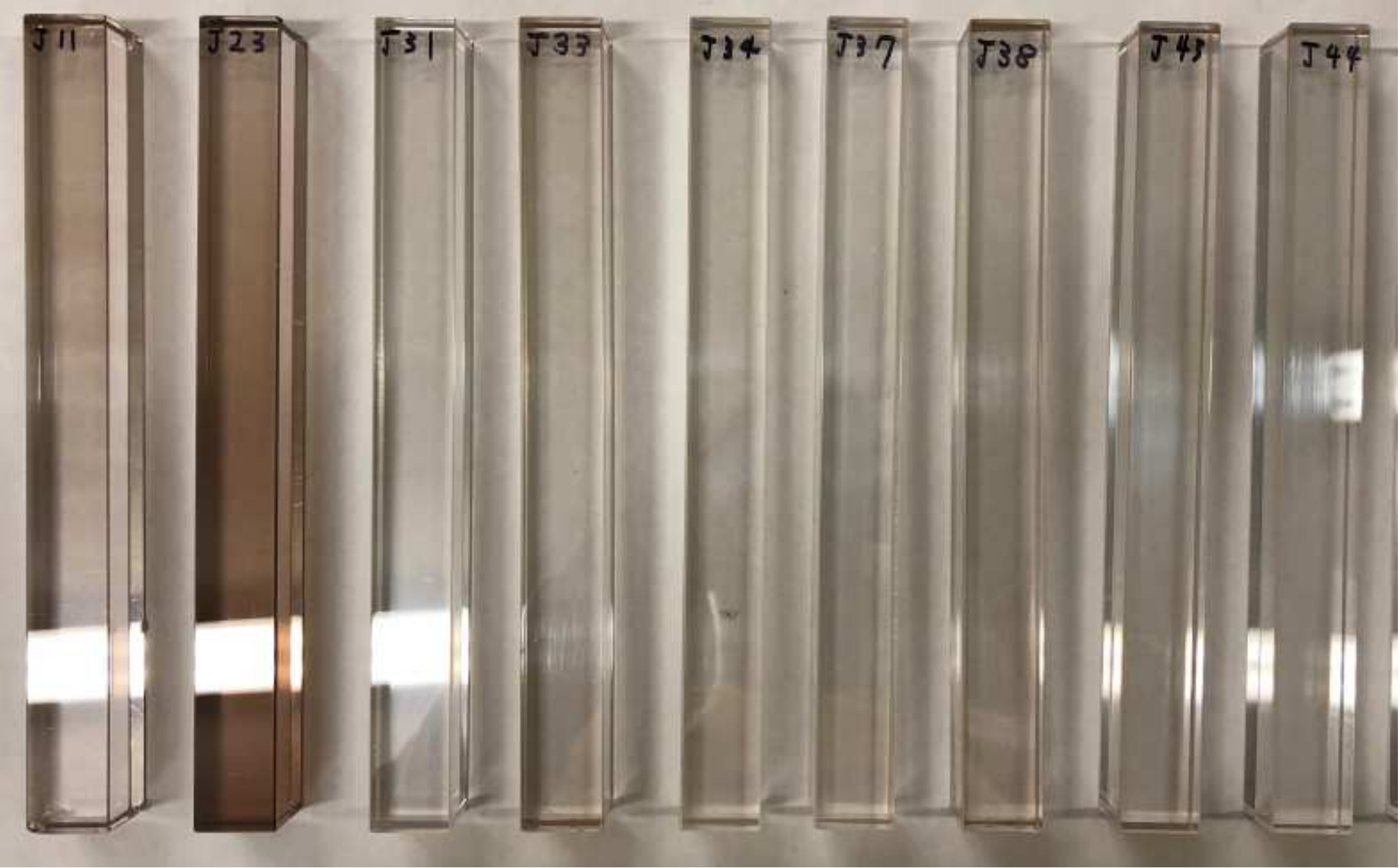} }
\caption{\label{fig:Radiation_Co60_visual} (Color online) Visual inspection of crystals after 30 Gy of radiation at 1 Gy/min}
\end{figure}

Figure~\ref{fig:Radiation_Co60_visual} illustrates the impact of an integral dose of 30 Gy at a dose rate of 1 Gy/min on a subset of 9 SICCAS samples. The radiation resistance varies considerably from sample to sample. While color center formation is significant in SIC-23 giving the sample a brown color, SIC-31 appear completely unaffected. 

\begin{figure}
\centering
% caption for subfigure a
\subfigure[\label{radiation-CO60-trans-good} ]
{\includegraphics[width=2.8in]{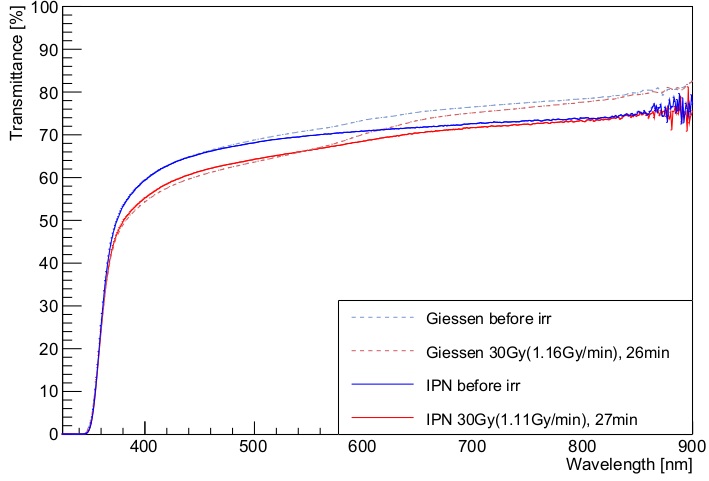} }
%caption for subfigure b
\subfigure[\label{radiation-CO60-trans-bad} ]
{\includegraphics[width=2.8in]{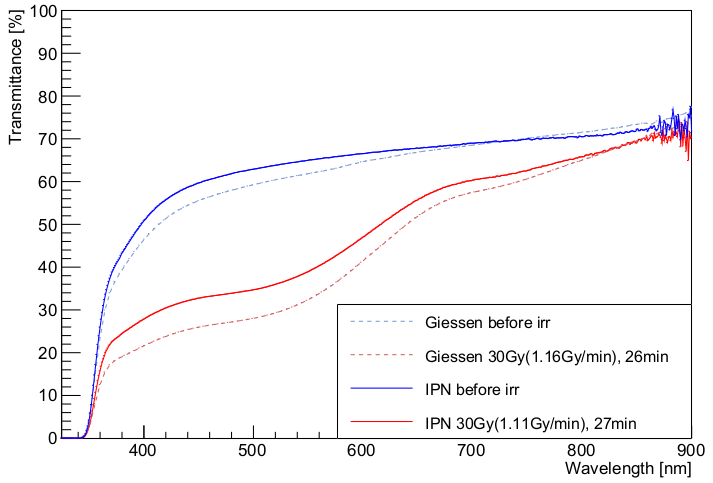}}
\caption{\label{fig:radiation-CO60-trans} (Color online) Transmittance after and before irradiation for a (a) good and (b) a bad crystal. The solid curves show measurements performed at Orsay and the dashed curves measurements performed at the Giessen facility.}
\end{figure}

The impact on transmittance can be seen in Fig.~\ref{fig:radiation-CO60-trans}. A sample of good radiation resistance has small variation in transmittance before and after irradiation. On the other hand, one observes significant radiation induced absorption throughout the spectrum, and in particular in the region $<$600nm for samples of poor radiation resistance. This absorption causes the yellow to brown coloring shown in Fig.~\ref{fig:Radiation_Co60_visual}. It should be noted that the shape of the radiation induced absorption varies from crystal to crystal.  

Radiation induced absorption results in significant degradation of the observed light yield. Samples showed saturation in their damage, which indicates the origin is most likely due to trace element impurities or defects in the crystal. The best samples show much less
degradation in light attenuation length and light output.

\subsection{Radiation induced absorption}

\begin{figure}
\centering
% caption for subfigure a
\subfigure[\label{radiation-CO60-abscoeff-good} ]
{\includegraphics[width=1.6in]{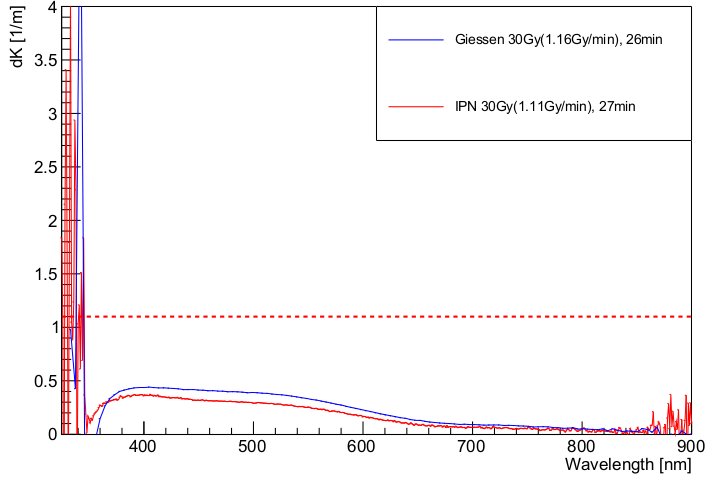} }
%caption for subfigure b
\subfigure[\label{radiation-CO60-abscoeff-bad} ]
{\includegraphics[width=1.6in]{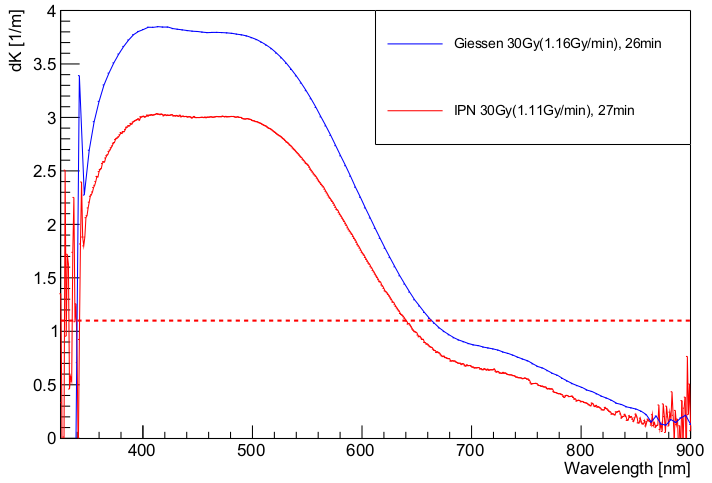}}
\caption{\label{fig:radiation-CO60-abscoeff} (Color online) Absorption coefficient for a (a) good and (b) a bad crystal.}
\end{figure}

Fig.~\ref{fig:radiation-CO60-abscoeff} shows the radiation induced absorption coefficient for crystal samples after a 30Gy dose of $^{60}Co$ $\gamma$ ray irradiation at at dose rate of 18Gy/min. The sample in Fig.~\ref{radiation-CO60-abscoeff-bad} shows significant radiation induced absorption.

%COMPARISON WITH CALTECH AND GIESSEN
Sample SIC-11 (significant scattering centers in bulk) was tested at the CUA, Caltech, Orsay, and Giessen facilities. The results agree within the uncertainty of the measurements. An illustration of the measurements at Orsay and Giessen is shown by the solid and dashed curves in Fig.~\ref{fig:radiation-CO60-trans}.

\subsection{Electron beam irradiation results}

The transmittance of some of crystals changed more than 15\% after an accumulated dose of 432 krad (at a dose rate of 1.3 Mrad/h), while others do not seem to show any effects of radiation damage. The change in transmittance for positions far from the front of crystals decreases with the distance. The effect of radiation damage is in part spontaneously recovered after a time period of 60 hours. Overall the results seem to suggest that the crystals can handle high doses at high dose rates. 

\begin{figure}
\centering
{\includegraphics[width=3.5in]{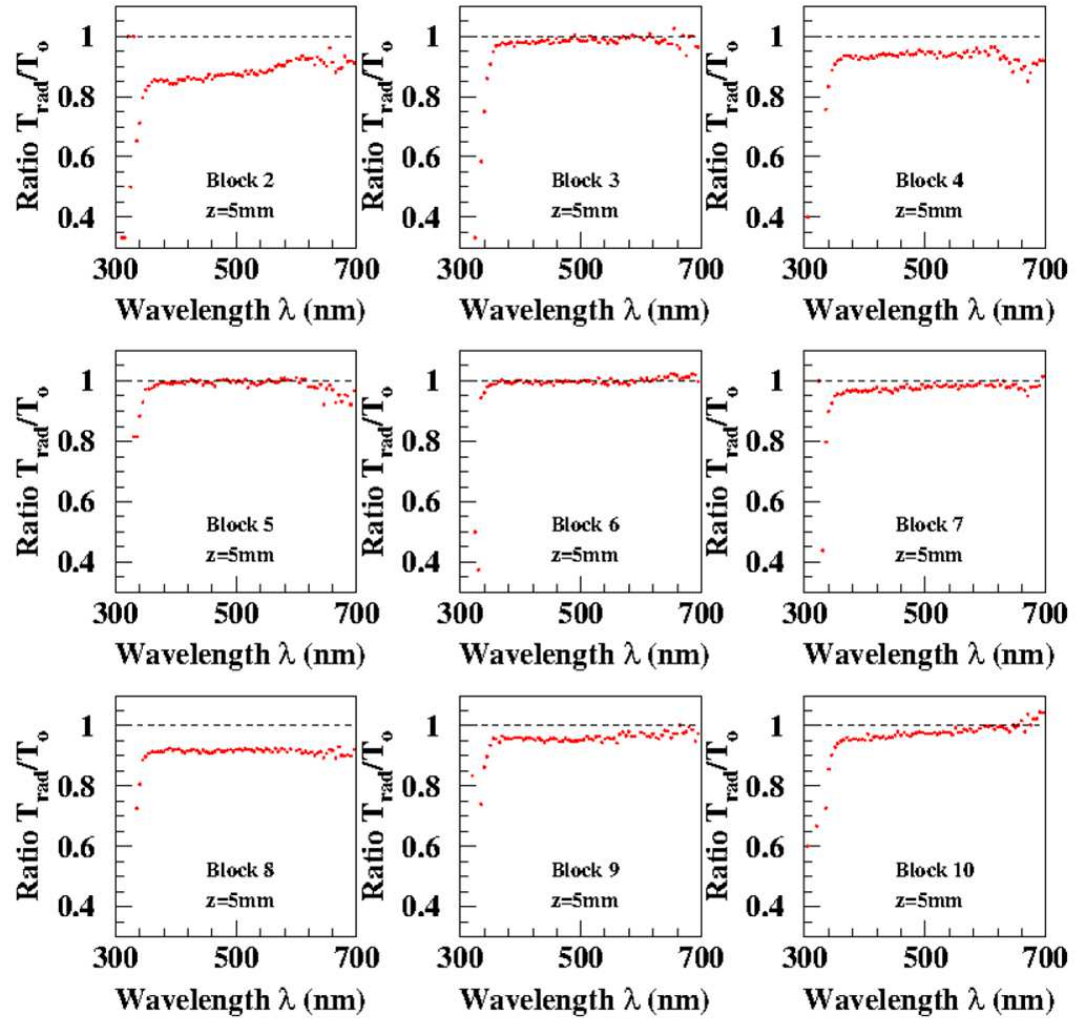} }
\caption{\label{fig:trans-degrad-432kr} (Color online) Transmission degradation of the PbWO$_4$ blocks after 432 krad accumulated dose at dose rates of 1.3 Mrad/h. Ratio of transmissions after and before irradiation reflects the level of crystal degradation. For example, crystal SIC-06 shown in the center panel was not damaged significantly.}
\end{figure}

One of the challenges in irradiation studies with beam is temperature control. Ideally one would control the temperature variation during the irradiation measurement within a few percent. This is difficult to achieve when working with an intense and narrowly focused beams, which give a high and concentrated dose to the crystals, and can even result in heating and thermal damage. As an example, for irradiation at a dose rate of 1.3 Mrad/hr, the temperature near the face of the crystal ramped up at a rate of 0.5 $^o$C/minute. For irradiation at a dose rate of 2.6 Mrad/hr, a rise of the temperature of more than 2 $^o$C/minute resulted in severe structural damage to the crystal after 
10 minutes. To reach higher doses crystals thus needed to be allowed to cool down between exposures. 

Another challenge in this measurement of radiation damage effects is to minimize surface effects. Ideally, one would measure the same spot before and after radiation minimizing surface effects in the path. Care was taken to ensure that this condition was satisfied and the flat distributions in Fig. \ref{fig:trans-degrad-432kr} seem to suggest that our setup satisfied this condition. To minimize the systematic uncertainty due to recovery of color centers with extremely fast times we carried out the transmittance measurement 10 minutes after irradiation.

\subsection{Thermal annealing and optical bleaching}

The radiation induced absorption can be reduced by thermal annealing, in which color centers are eliminated by heating the crystal to a high temperature, or optical bleaching, in which light is injected into crystals. Color center annihilation is wave length dependent. Thermal annealing is beneficial to recover individual or small numbers of crystals. In a medium to large detector like the NPS optical bleaching is the preferred method. 

{\subsubsection{Thermal Annealing}

Thermal annealing was done at 200$^\circ$C for 10 hours. The protocol included a ramp up/down procedure at 18$^\circ$C per hour starting/ending at room temperature. The temperature profile used to anneal the crystals is shown in Fig.~\ref{fig:thermal-anneal-profile}. The transmittance of crystals exposed to an integrated dose of 30 Gy EM radiation is shown in Fig.~\ref{fig:radiation-CO60-trans}. For crystals received from the vendors and not exposed to radiation no significant differences in optical properties were found before and after thermal annealing.

\begin{figure}
\centering
{\includegraphics[width=3.0in]{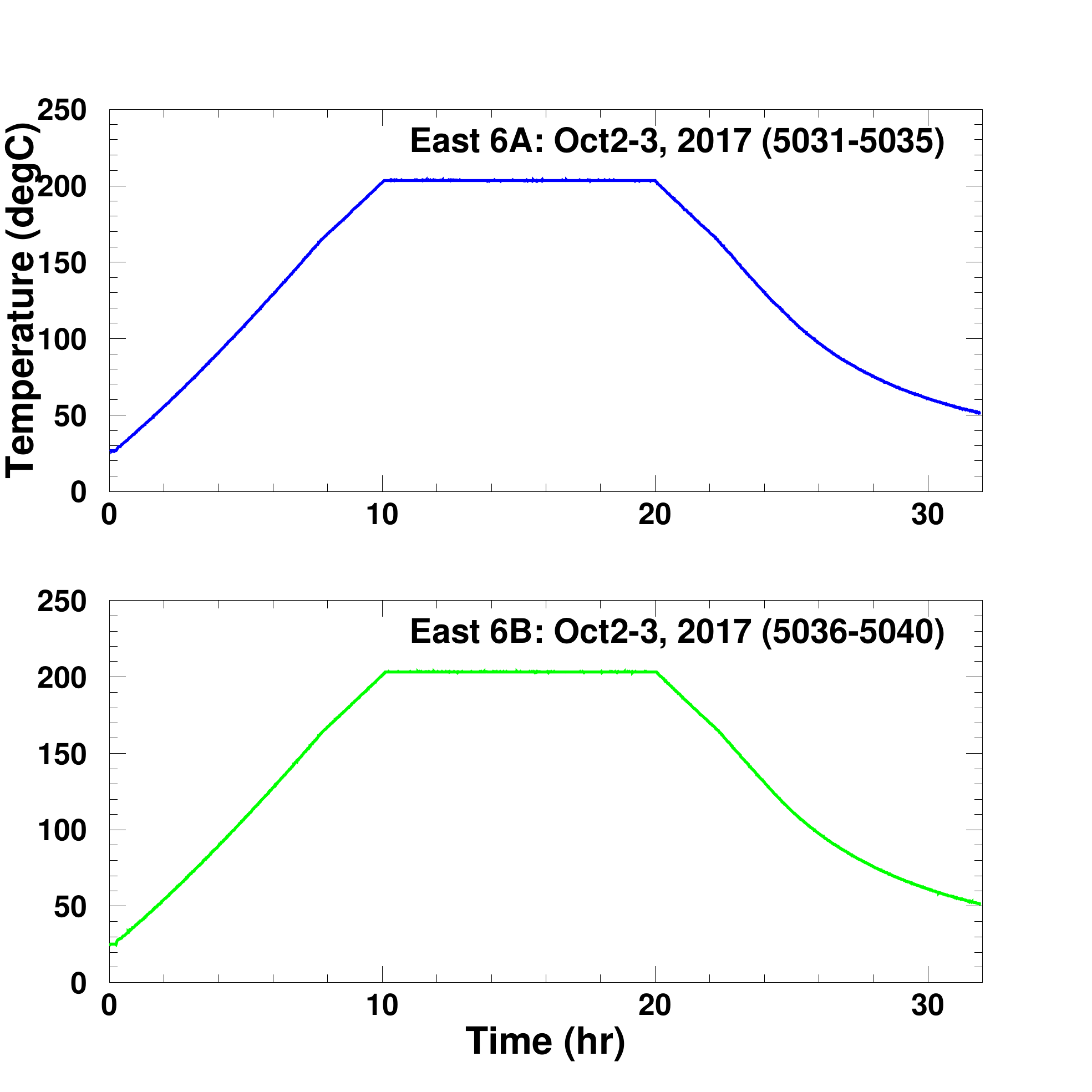} }
\caption{\label{fig:thermal-anneal-profile} (Color online) Temperature profiles for two of the furnaces used for thermal annealing of the crystals.}
\end{figure}

%GRAPH OF IRRADIATED CRYSTAL BEFORE AND AFTER THERMAL ANNEALING

%Two methods of optical bleaching to recover crystal irradiation damage have been studied. One is based on Blue (or UV), the other on infra-red (IR) light. 

{\subsubsection{Optical Bleaching}

Studies show that with blue (UV) light of wavelength $\lambda\sim$400-700 nm~\cite{Novotny-2012}, nearly 90\% of the original amplitude can be restored within 200 minutes with photon flux of $\sim 10^{16}$ photon/s. Light of short wavelength is most effective for recovery, but recovery at longer wavelength (700-1000 nm) recovery is also possible. It works very well for low doses ($\sim$3 krad), but its efficiency compared to blue light is reduced by a factor of $\sim$20-50. This can be compensated by using high intensity IR light ($\geq 10^{16}$ photons/s per block). Studies show that at dose rates $\sim$1 krad/h with a IR light of $\lambda\geq$900 nm and intensity $\sim 10^{16}-10^{17}~\gamma$/sec one may continuously recover degradation of the crystal~\cite{Novotny-2012,Novotny-2009}. 
Fig.~\ref{fig:uv-recovery-time} illustrates the effect of blue light and IR curing on 2 crystal samples (one with low, one with high radiation resistance) from SICCAS. The effect of either type of curing is similar for the crystal with good radiation resistance, whereas the blue light curing results in faster recovery for the crystal with low radiation resistance. 

\begin{figure}
\centering
{\includegraphics[width=3.5in]{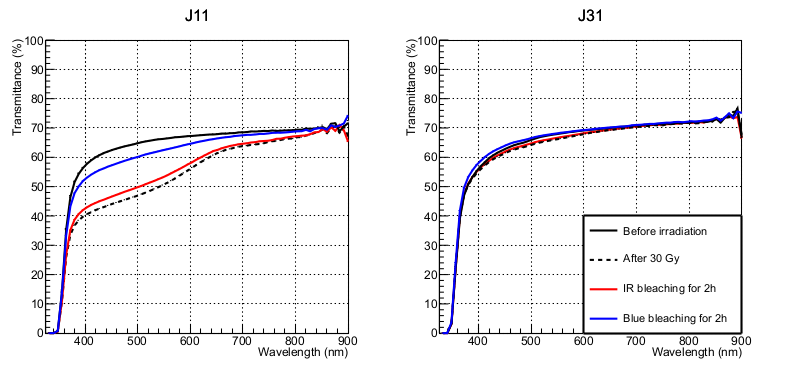} }
\caption{\label{fig:uv-recovery-time} (Color online) Impact of blue light and IR curing on 2 crystal samples with low (left) and high (right) radiation resistance. The black solid and black dashed curves denote the transmittance of the crystal before and after 30 Gy radiation dose, respectively. The blue curve shows the transmittance after 2 hours of blue light curing, the red curve the transmittance after 2 hours of IR curing.}
\end{figure}

An advantage of IR curing is that it can in principle be performed continuously, even 
without turning off the high voltage on the PMTs as long as the IR light is out of the PMTs quantum efficiency region. To test this assumption the emission intensity of the Infrared LED LD-274-3 and TSAL7400 versus driving current were been measured. The peak wavelengths are 950 nm for LD-274-3 and 940 nm for TSAL7400.

The LEDs were mounted on a special support structure and the intensity of the emitted light was measured with a calibrated photodiode (S2281) with an effective area of 100 mm$^2$. The distance between LED and photodiode was variable from 0.5 cm to 20 cm. The photodiode dark current when the LED was off was on the level of $\sim$0.001 nA. The emitted light was measured with a PMT (Hamamatsu R4125) installed at the front of the LED. The measurements were done at different LED driving currents (from 0 up to 100 mA), at distances 0.5 cm, 3cm, and 16 cm (18 cm), with and without a PbWO$_4$ crystal attached to the PMT. To eliminate contamination of short wavelength light in the emission spectrum of the IR LEDs  measurements were made with and without a 900 nm long-pass filter. 

Our results show that the Hamamatsu R4125 has a very low, but not negligible sensitivity to infrared light. Since even a low quantum efficiency may reduce the PMT live time for a typical IR curing flux of $N\sim10^{16}-10^{17}~\gamma/sec$ and because of the lower efficiency relative to blue light (see Fig.~\ref{fig:uv-recovery-time}, the NPS optical bleaching system is based on blue (UV) light.

\section{Structural and chemical analysis}
\label{sec-structural-analysis}

The chemical composition of the crystals were investigated at the Vitreous State Laboratory (VSL) using a combination of standard chemical analysis methods including XRay Fluorescence (XRF) and ICP-MS. The surface analysis was performed with a scanning electron microscope with EDS and WDS systems and nanomanipulator (JEOL 6300, JEOL 5910).

\subsection{Surface Properties}
\label{subsec:Surface-property}

Figure~\ref{fig:crystals-surface} shows the surface quality of representative crystals from Crytur at 50 $\mu$m and SICCAS at 500 $\mu$m. For comparison, a BTCP sample was analyzed as well. The surface of the Crytur crystal is well-polished with negligible mechanical flaws. The SICCAS crystal has long scratches on the surface and also other flaws as shown. The BTCP crystal surface has scratches, which is expected as this crystal had been shipped multiple times without re-polishing. 

\begin{figure*}
\centering
% caption for subfigure a
\subfigure[\label{crytur-surface} Crytur]
{\includegraphics[width=1.6in]{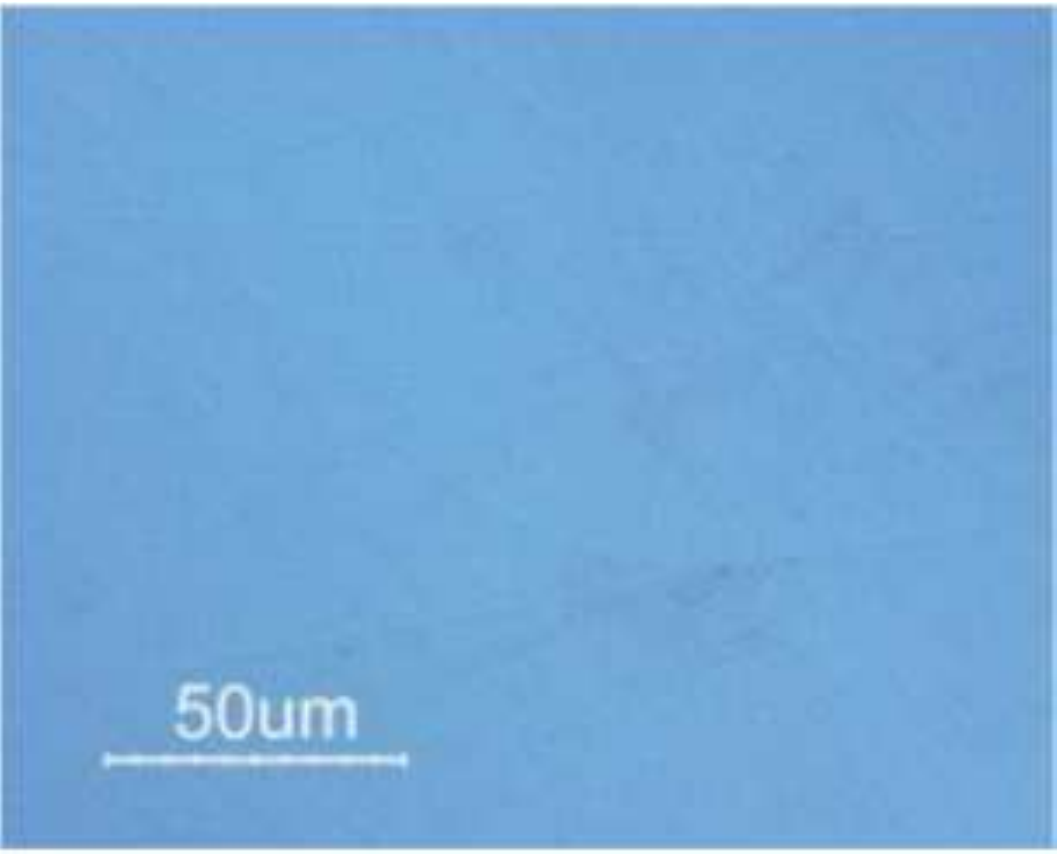} }
%caption for subfigure b
\subfigure[\label{btcp-surface} BTCP]
{\includegraphics[width=1.6in]{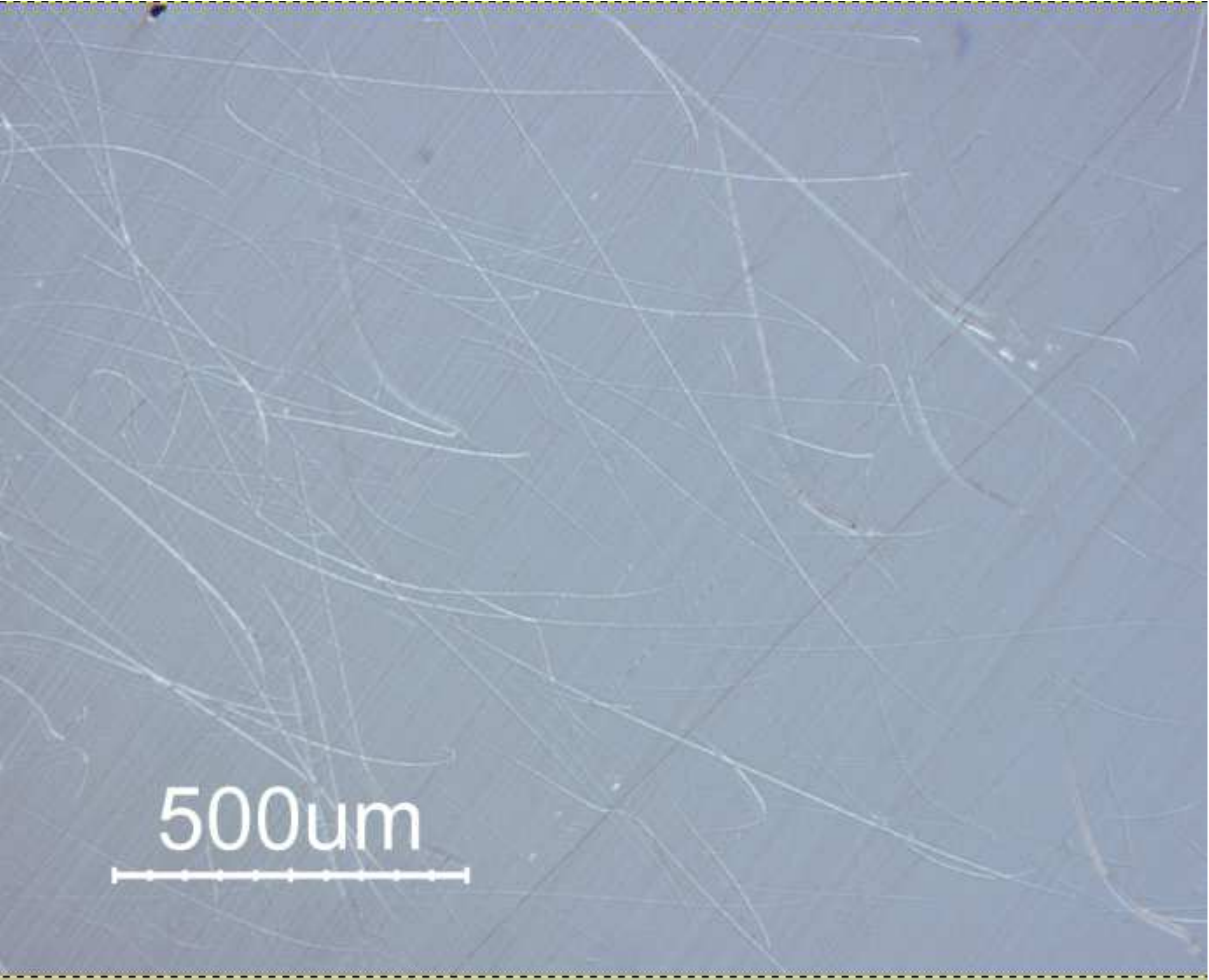}}
%caption for subfigure c
\subfigure[\label{siccas-surface} SICCAS]
{\includegraphics[width=1.6in]{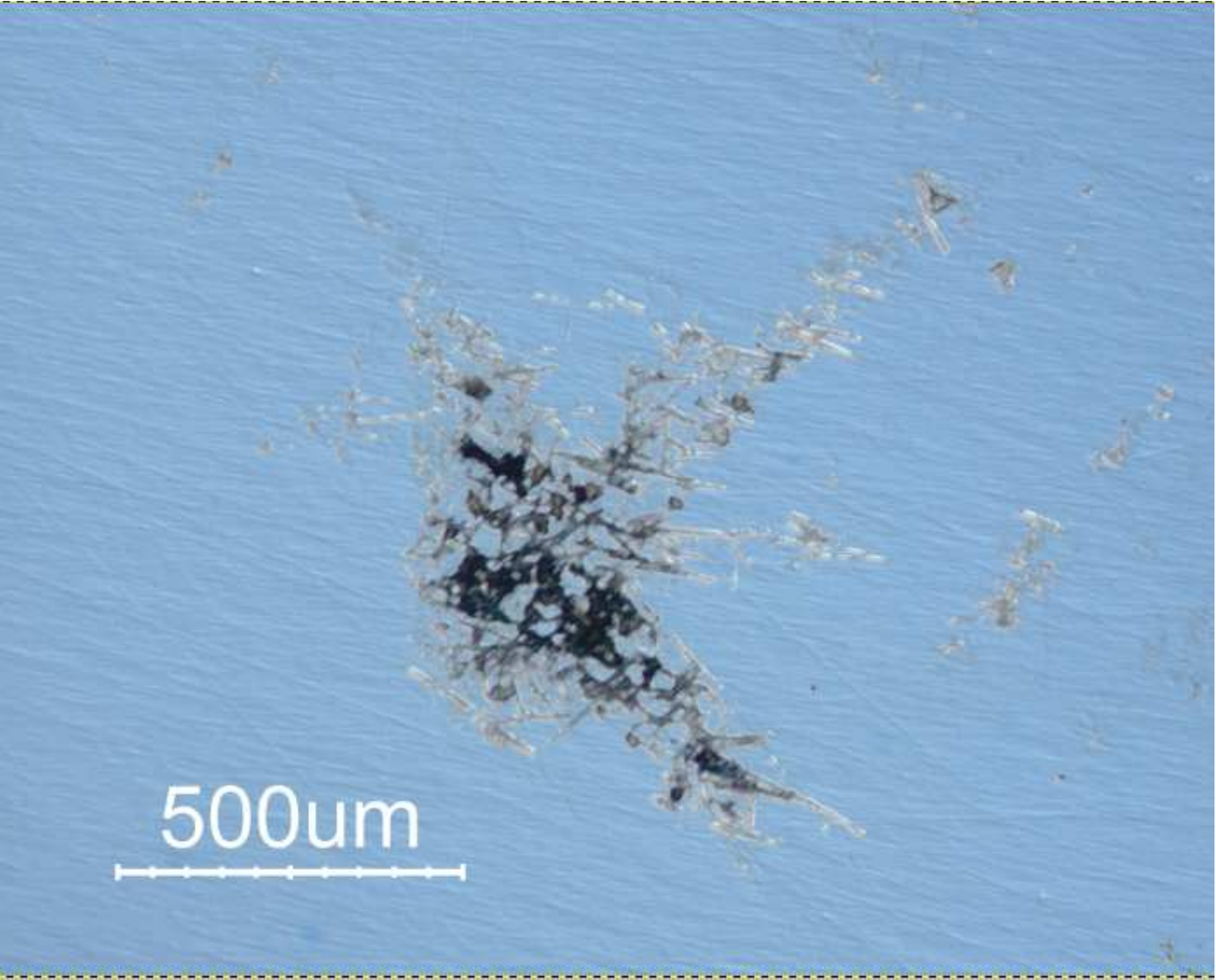}}
\caption{\label{fig:crystals-surface} (Color online) Microscope surface 
analysis of PbWO$_4$ crystals from Crytur (a), BTCP (b) 
and SICCAS produced in 2017 (c).}
\end{figure*}

Looking even deeper into the crystal defects of the SICCAS samples (see Fig.~\ref{fig:SICCAS-crystals-defects}) reveals  bubbles and deep pits up to 20 $\mu$m inside the bulk. The size of these bubbles can be on the order of 100 $\mu$m. These flaws can be correlated with an observed very high, but position dependent light yield inducing non-uniformities, as well as a very low transmittance around 400-450 nm.  

\begin{figure*}
\centering
% caption for subfigure a
\subfigure[\label{crytur-surface} SICCAS crystals bubbles]
{\includegraphics[width=1.6in]{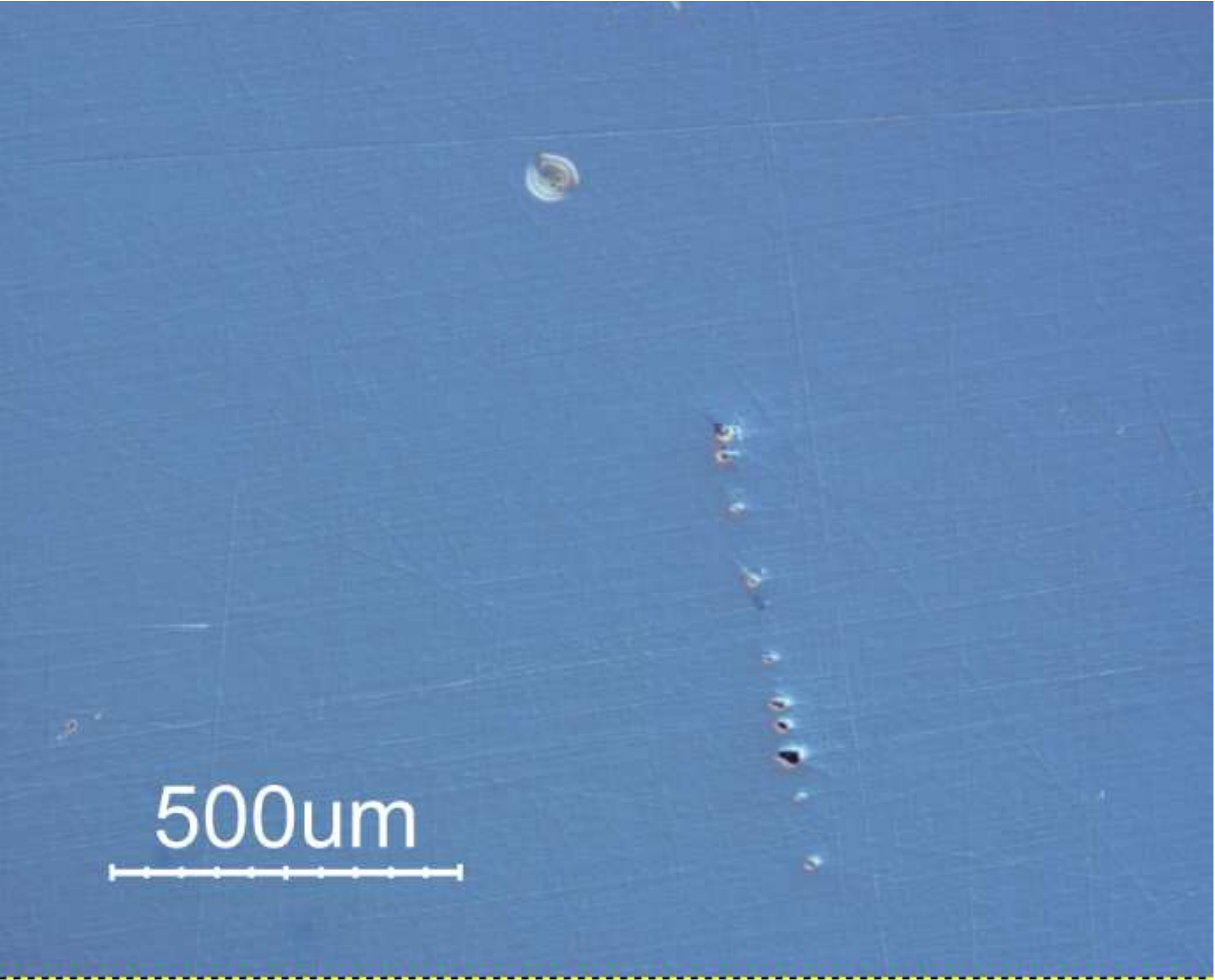} }
%caption for subfigure b
\subfigure[\label{btcp-surface} SICCAS crystals scratches]
{\includegraphics[width=1.6in]{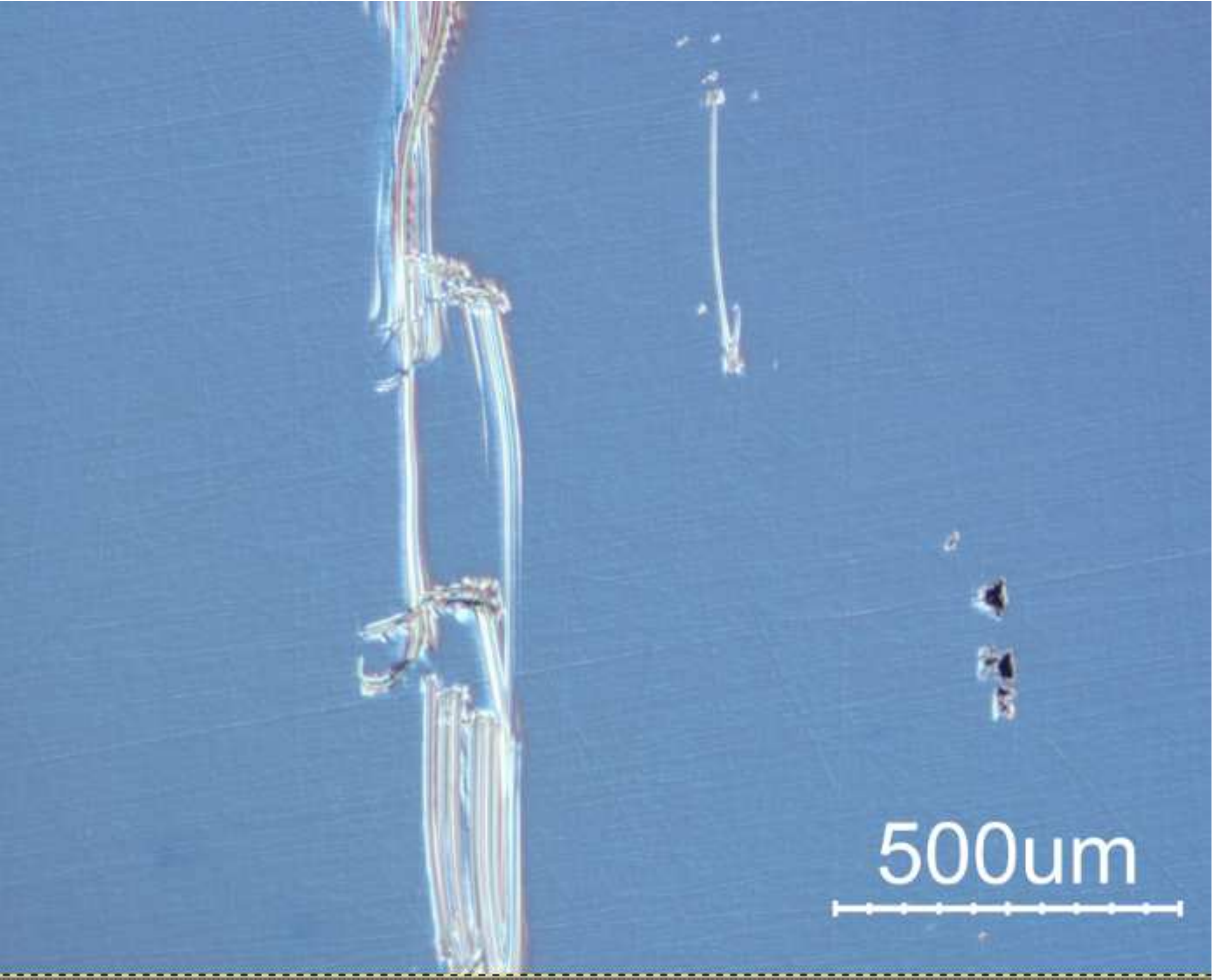}}
%caption for subfigure c
\subfigure[\label{siccas-surface} SICCAS crystals pits]
{\includegraphics[width=1.6in]{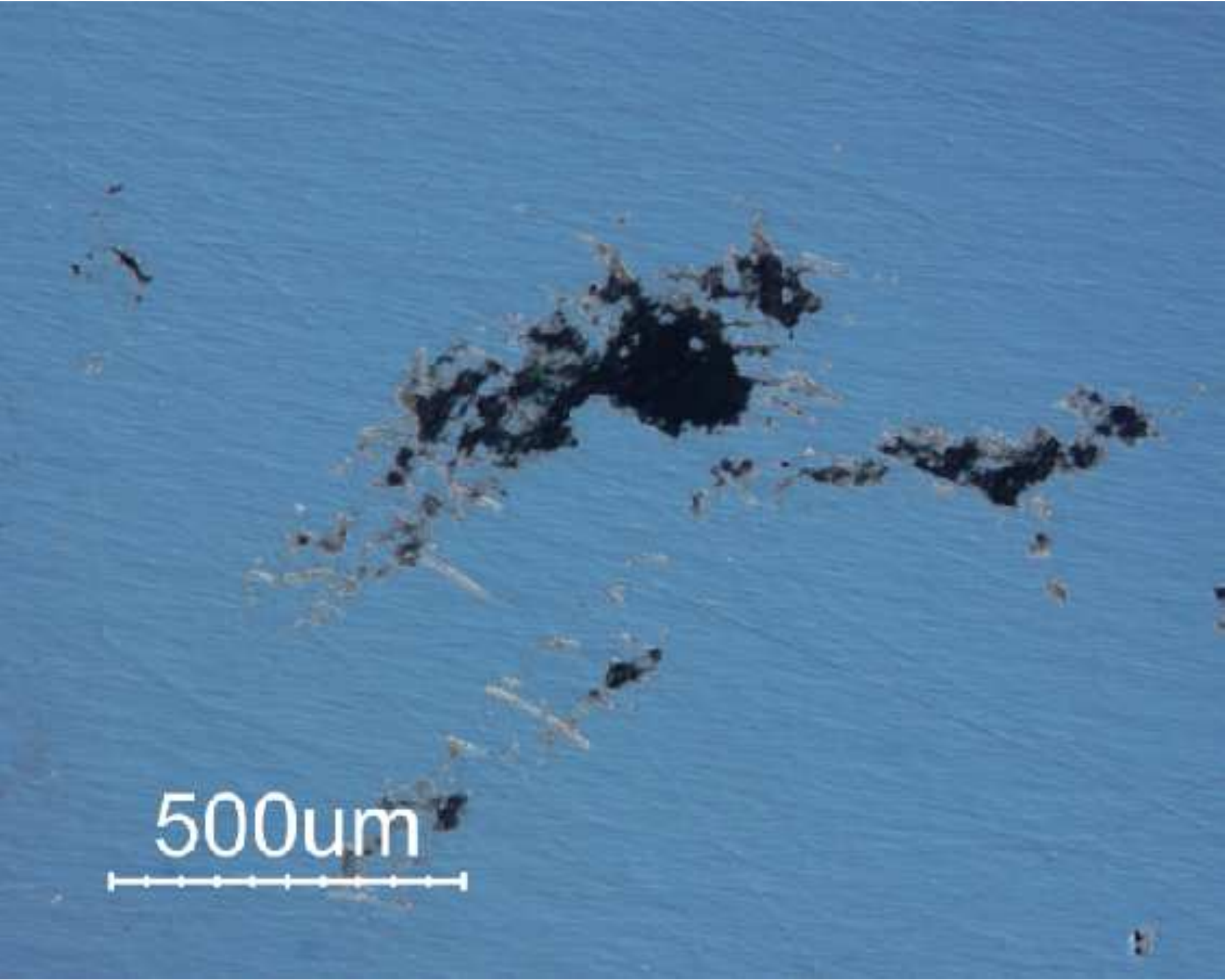}}
\caption{\label{fig:SICCAS-crystals-defects} (Color online) Microscope images 
of bubbles (a), deep scratches (b) and pits (c) observed in SICCAS crystals 
produced in 2017.}
\end{figure*}

\subsection{Chemical composition analysis}
\label{subsec:chem-composition}

%\subsubsection{XRF studies}

Real crystals contain large numbers of defects, ranging from variable amounts 
of impurities to missing or misplaced atoms or ions. It is impossible to obtain
 any substance in 100\% pure form. Some impurities are always present.
Even if a substance were 100\% pure, forming a perfect crystal would require 
cooling infinitely slowly to allow all atoms, ions, or molecules to find their 
proper positions. Cooling usually results in defects in crystals.
In addition, applying an external stress to a crystal (cutting, polishing) may 
cause imperfect alignment of some regions of with respect to the rest.
In this section, we discuss how chemical composition can impact some of the crystal properties. 

Samples on the order of 100 microgram were taken from each crystal using a 
method developed by the VSL. The method is non-destructive and does not impact 
the crystal optical properties. The latter was verified with dedicated measurements, e.g. of the light yield before and after the sample was taken. Approximately 10-15\% of the crystals were investigated in this study. 

Figure \ref{fig:SICCAS-crystals-composition} shows a general overview of the variation in composition for a subset of 15 randomly selected SICCAS crystals in terms of the element oxides. Also shown are the results for two randomly selected CRYTUR (column 4, 5) and one BTCP crystal (column 18). The two major materials (PbO and WO$_3$) used in crystal growing are not shown. The variation in these materials among all crystals is small (0.5-0.7\% on average), which one might interpret as differences in optical properties being due to other contributions in the chemical composition (see results of statistical analyses in the next paragraphs) or mechanical features. Crystals that pass all optical specifications seem to have a noticeable contribution from iron oxide and smaller contributions from at most two other oxides. On the other hand, crystals that fail all or a large number of optical specifications have at least three contributions other than iron. 

\begin{figure}
\centering
{\includegraphics[width=3.5in]{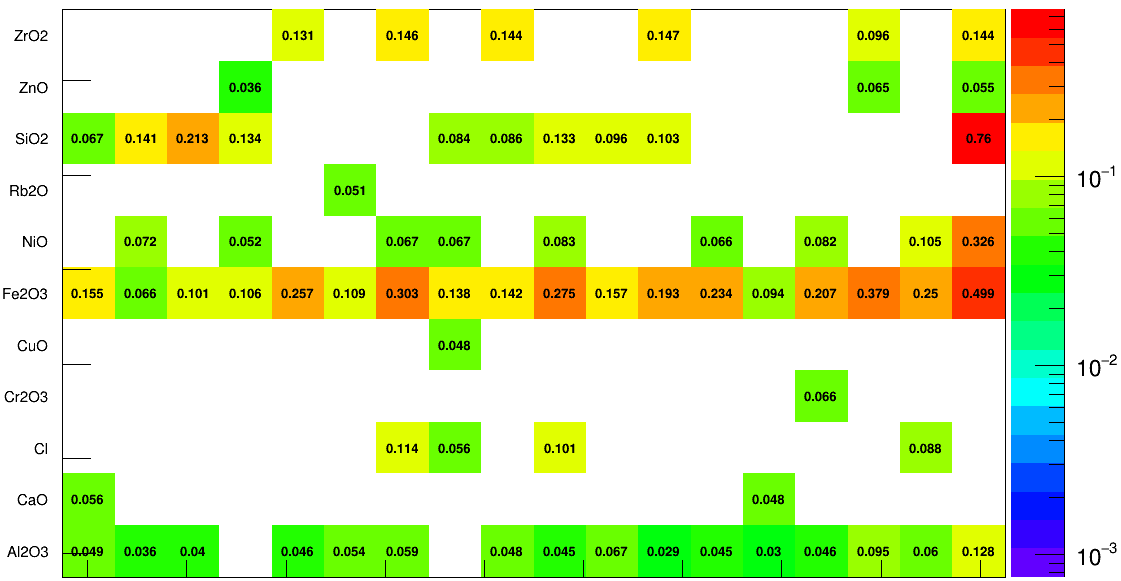} }
\caption{\label{fig:SICCAS-crystals-composition} (Color online) Crystal composition from XRF analysis. The two major materials (PbO and WO$_3$) used in PbWO$_4$ crystal growth are not shown.}
\end{figure}

To investigate the importance of the variation in lead and tungsten oxides, as 
well as those of the other elements observed in chemical composition analysis, 
statistical analyses were carried out. The first method is a multivariate 
approach in which correlations are estimated by a pairwise method. The results 
are shown in Fig.~\ref{fig:crystal-composition}. A clear dependence of the 
optical transmittance on the stoichiometry of lead and tungsten oxides can be 
seen. The light yield does not seem to depend on this stoichiometry.

\begin{figure}
\centering
{\includegraphics[width=3.5in]{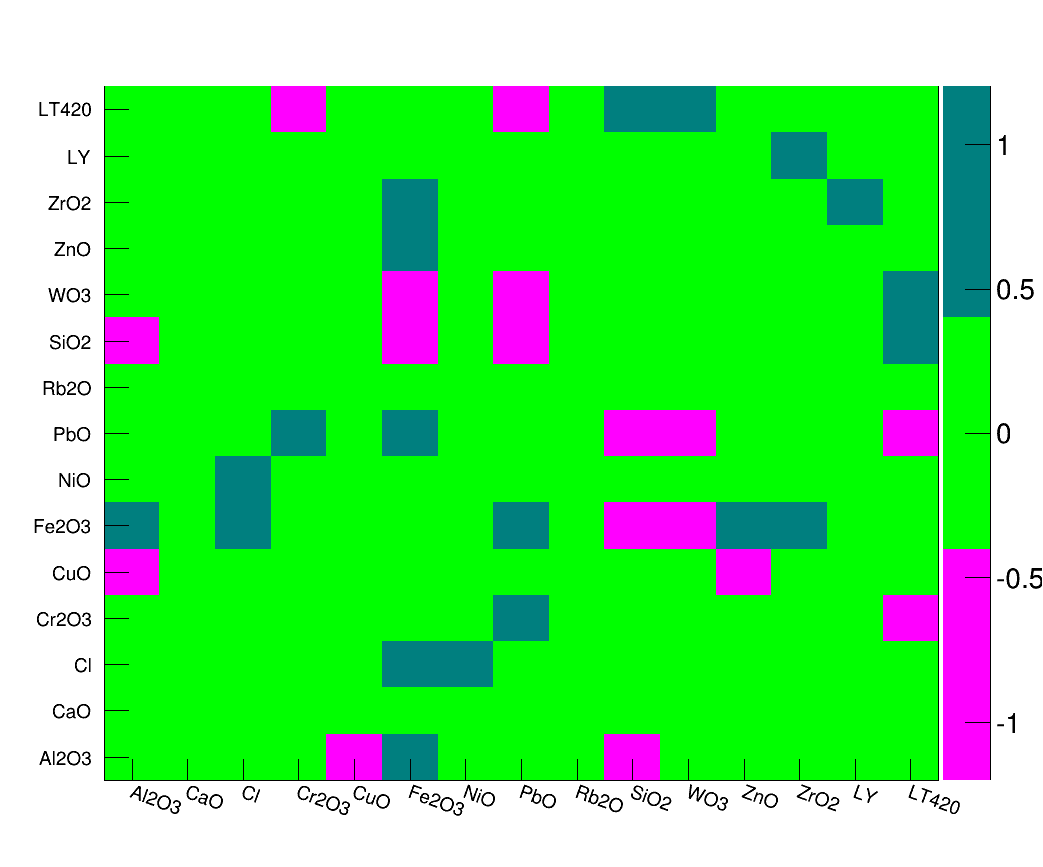} }
\caption{\label{fig:crystal-composition} (Color online) Multivariate analysis results. A clear
dependence of optical transmittance on PbO/WO$_3$ stoichometry can be observed. 
Light yield appears independent on it.}
\end{figure}

The second statistical method uses partial least squares to construct two 
correlation models and assess effects of individual variables. The results for 
two resulting models assessing the impact of chemical composition on light 
yield and optical transmittance is shown in 
Figure \ref{fig:LY-and-OT-vs-composition}. Zr, Ni, and Ca seem to be
 most relevant for light yield, while Si and to a lesser extent Cr seem most 
relevant for transmittance at 420 nm. 

\begin{figure}
\centering
% caption for subfigure a
\subfigure[\label{light-yield-vs-composition} Light yield vs composition]
{\includegraphics[width=1.6in]{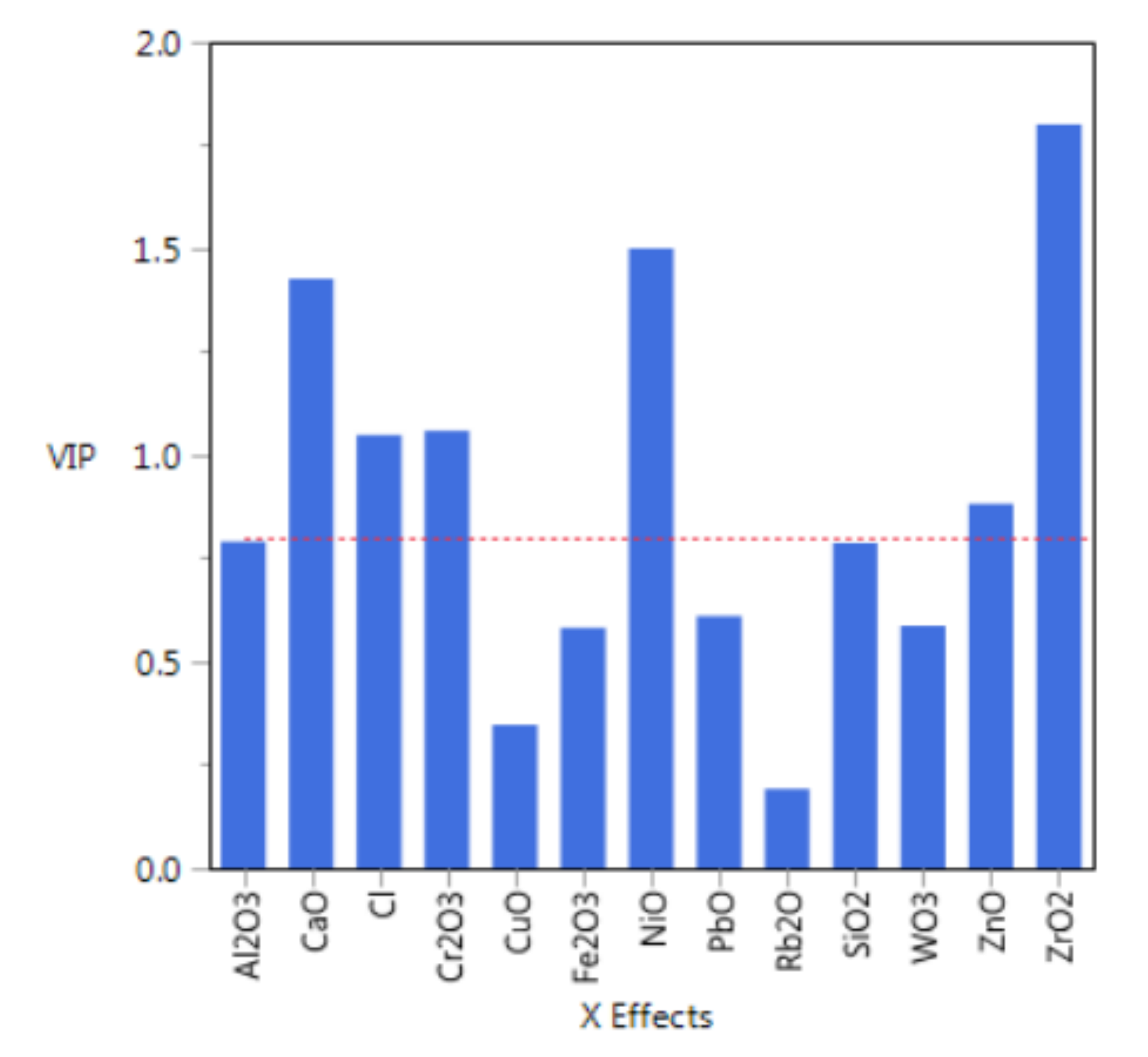} }
%caption for subfigure b
\subfigure[\label{transmittance-vs-composition} Transmittance vs composition]
{\includegraphics[width=1.6in]{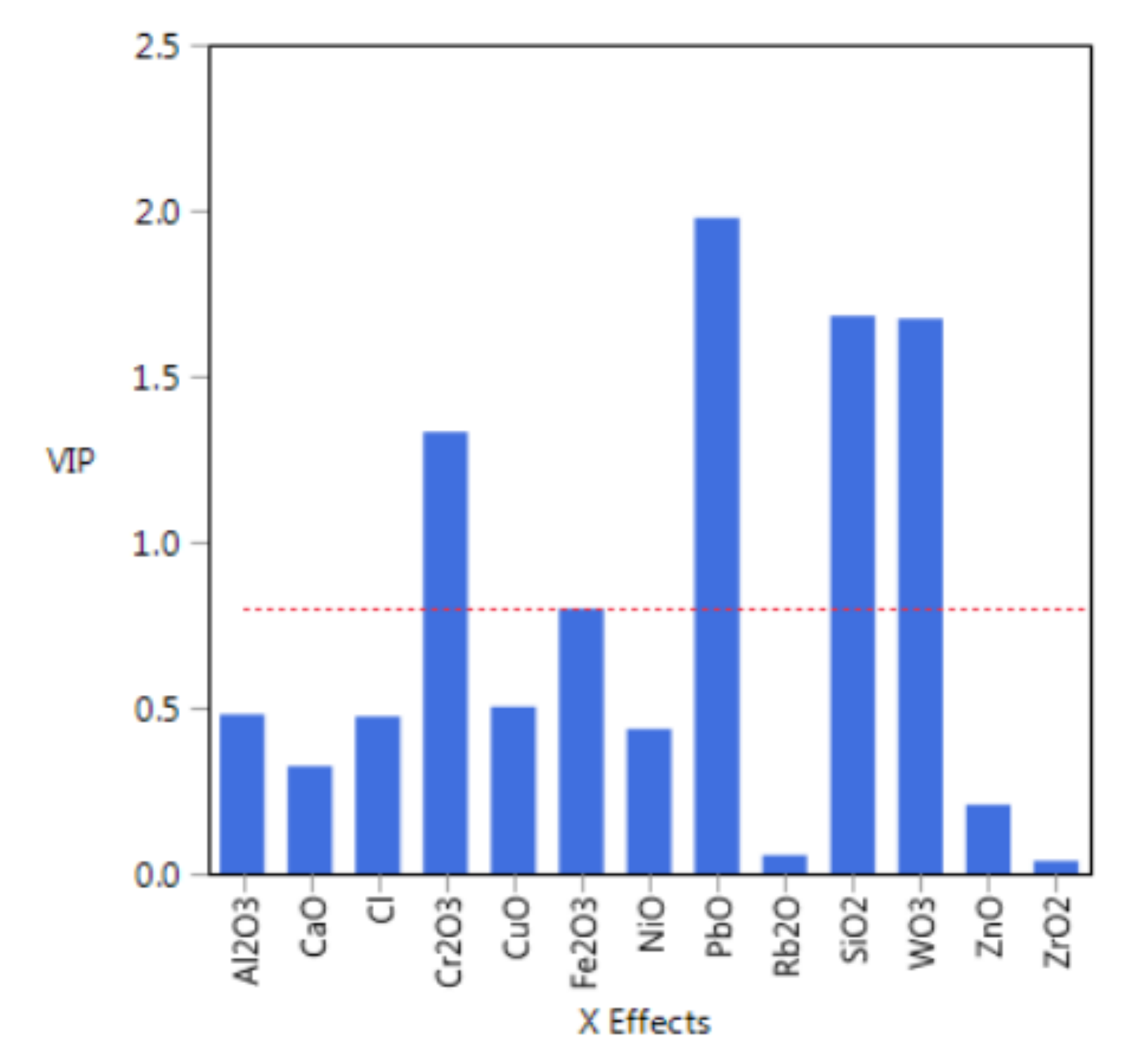}}
\caption{\label{fig:LY-and-OT-vs-composition} (Color online) Effect of 
individual elements of chemical composition on light yield (a) and optical 
transmittance (b) based on a partial least square statistical analysis.}
\end{figure}

\section{Beam test program with prototype}
\label{sec-prototype}

\begin{figure}[h!]
\centering
\includegraphics[width=3.0in]{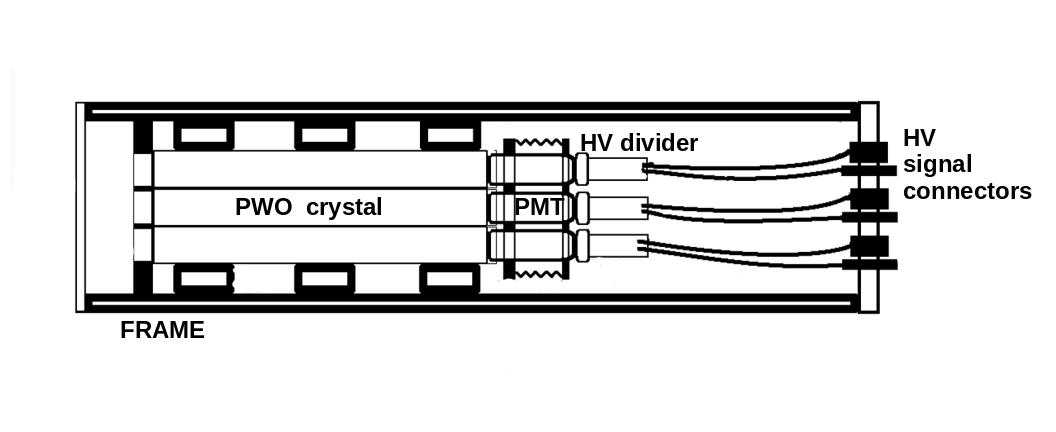}
\caption{(Color online) Neutral Particle Spectrometer
 (NPS) prototype schematic view.}
\label{fig:nps-prototype1}       % Give a unique label
\end{figure}

A first prototype was constructed at JLab using 3D printing technology. Fig. \ref{fig:nps-prototype1} shows a schematic view of the prototype mechanical structure. The prototype consists of a 3x3 matrix of PWO crystals, placed inside a brass box. The stack of crystals is  fixed to the box using  3D-printed plastic holders. The front face of the prototype box is covered with a 2 mm thick plastic plate. The plastic mesh plate is placed in front of the crystal stack and is mounted to the prototype frame to prevent individual crystals from sliding in the forward direction. The crystals are wrapped with  an 65 $\mu$m ESR reflector and a 30 $\mu$m thick Tedlar film to provide light tightness. Each crystal is coupled to a R4125-01 Hamamatsu PMT using an optical grease. The PMTs are attached to the crystals using two plastic holder plates. The front plate is attached to the side wall of the prototype frame and has nine holes allowing the PMT`s to slide in the forward direction towards crystals. The movable back PMT plate holds the PMTs and provides pressure needed for optical coupling using springs, which are connected between the plates in each corner. The back plate has holes for PMT pins, to attach dividers. Each PMT is powered and read out using a HV divider with an integrated preamplifier designed at Jefferson Lab. High voltage and signal cables are connected to the SHV and LEMO connectors installed in the back plate of the prototype box.

Performance of the calorimeter prototype was studied using secondary electrons provided by the Hall D Pair Spectrometer (PS)\cite{Pair_spectrometer}. The schematic view of the Pair Spectrometer is presented in Fig. \ref{fig:nps-prototype2} Electron-positron pairs are created by beam photons in a 750 $\mu$m Beryllium converter. The produced leptons are deflected in a 1.5 T dipole magnet and are detected using two layers of scintillator counters positioned symmetrically with respect to the photon beam line. In each arm, there are 8 coarse counters and 145 high-granularity counters. The coarse counters are used in the trigger. The high-granularity hodoscope is used to measure the lepton momentum; the position of each counter corresponds to the specific energy of the deflected lepton. Each detector arm covers a momentum range of e ± between 3.0 GeV/c and 6.2 GeV/c. The energy resolution of the pair spectrometer is estimated to be better than 0.6\%. The calorimeter prototype was positioned behind the PS as shown in Fig.  \ref{fig:nps-prototype2} The energy of electrons passing through the center of the middle module was measured using the PS hodoscope and corresponded to 4.7 GeV. High voltages for nine prototype channels were provided by CAEN A1535SN module. Signals from PMTs are digitized using a twelve-bit 16 channel flash ADC operated at 250 MHz sampling rate \cite{FADC-250}. Digitized amplitudes are integrated in a time window of 68 ns. Readout of the prototype was integrated to the global GlueX DAQ system. Data were collected in parallel with the GlueX \cite{GlueX} using the pair spectrometer trigger, which was produced by the electron-positron pair and is required for the luminosity determination in GlueX.

\begin{figure}[h!]
\centering
\includegraphics[width=3.0in]{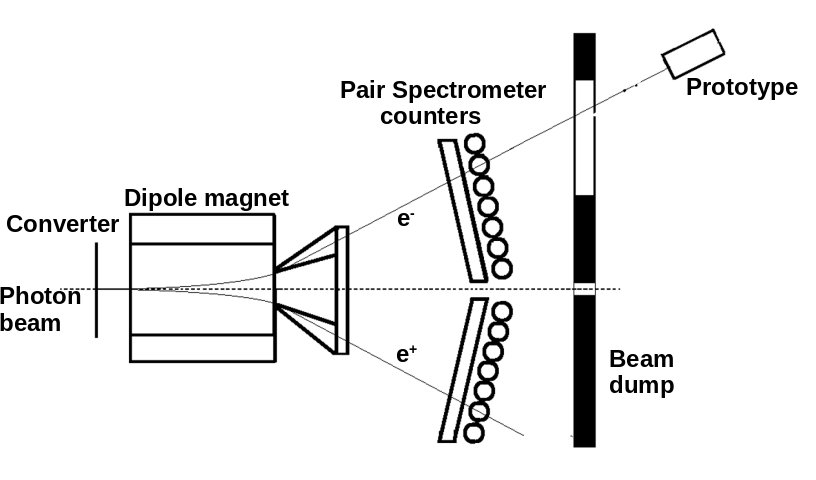}
\caption{(Color online) Position of the calorimeter behind the HallD Pair Spectrometer.}
\label{fig:nps-prototype2}       % Give a unique label
\end{figure}

We calibrated the energy response (gain factors) of each calorimeter module using two independent methods:
\begin{itemize}
\item Direct energy calibration. Three modules in each row were calibrated by measuring energy depositions (in units of fadc counts) for electrons incident on the middle of each cell. Modules from other rows were subsequently calibrated by lowering and lifting the prototype by ±2 cm (the module size) and exposing corresponding rows to the beam. 
\item Using regression calibration. Calibration coefficients were obtained by minimizing the difference between the total energy deposited in the 3x3 calorimeter prototype and the electron energy reconstructed by the Pair Spectrometer. The calibration was performed for events where electrons hit the center of the middle module:
\begin{equation}
\sum_{events} (\sum_{i=1}^{Nseg}k_i A_i - E_{ps})^2 \rightarrow min
\end{equation}
where Nseg  is the number of modules in the cluster, k  is the calibration coefficient,  A is the signal pulse integral, and Eps  is the  electron energy measured by the pair spectrometer.
\end{itemize}
These two calibration methods provided consistent results.
Fig. \ref{fig:nps-prototype3} a) and b) show reconstructed energy in the 3x3 calorimeter for ~4.7 GeV electrons incident on the middle of the central module. The calorimeter was constructed using CRYTUR and SICCAS crystals and was tested during the spring run of 2019. The measured resolution was ~1.6\% and ~1.5\% for CRYTUR, SICCAS crystals, respectively. We also observed about 6\%  larger light yield for SICCAS crystals, which can potentially account for slightly better energy resolution. Our results show that beam tests with the 3x3 calorimeter provide a method for quick configuration tests, estimations of energy resolution, and comparison of crystal properties. We also constructed a 12x12 prototype calorimeter that allowed us to take data over a larger energy range and also to study linearity, e.g., of the high voltage divider and amplifier. The results from this beam test will be published in a forthcoming publication~\cite{Comcal}.

\begin{figure}[!htbp]
\centering
\subfigure[\label{fig:nps-prototype3_a} SICCAS crystals]
% caption for subfigure a
{\includegraphics[width=1.6in]{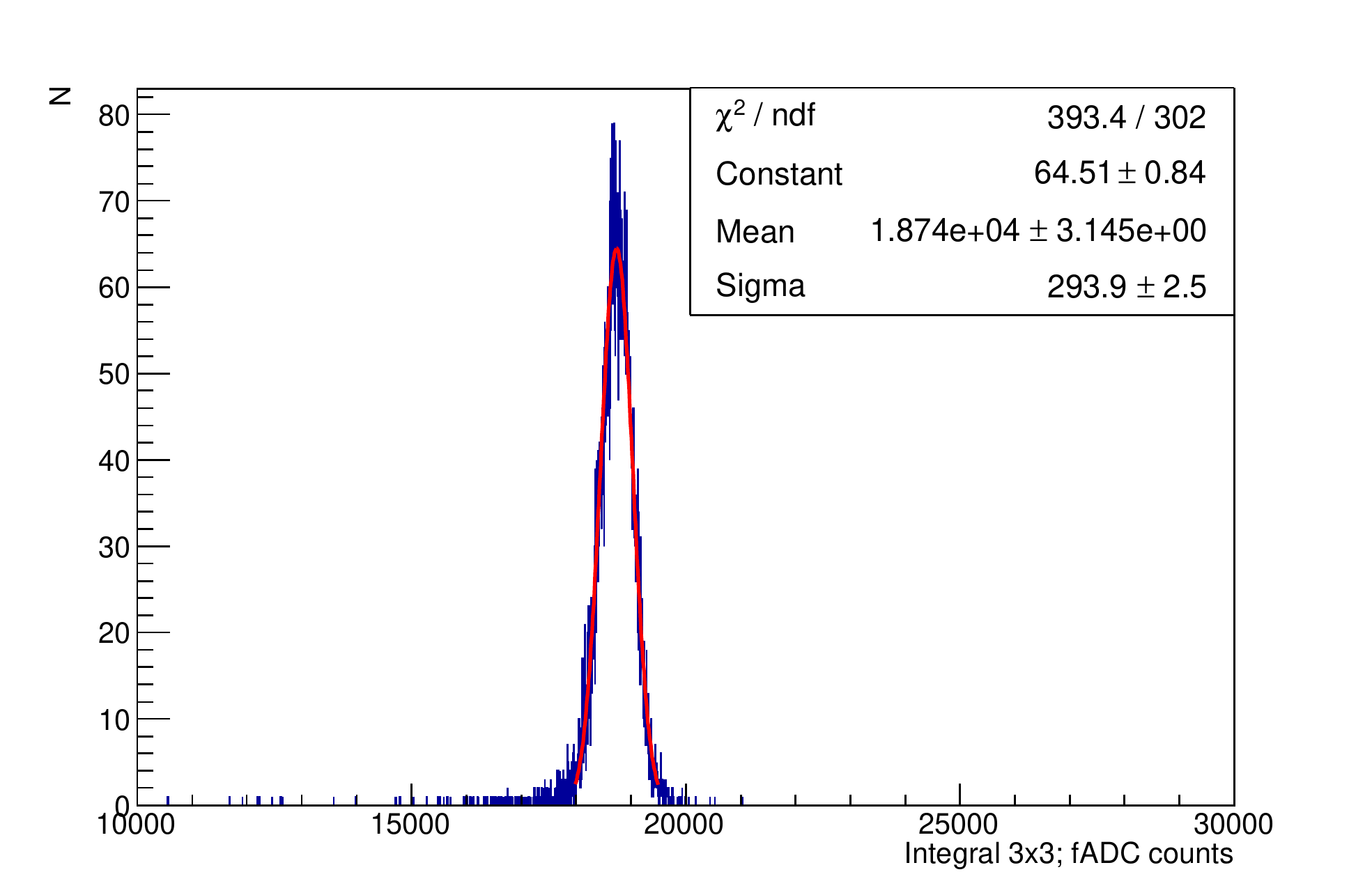}}
\subfigure[\label{fig:nps-prototype3_b} CRYTUR crystals]
%caption for subfigure b
{\includegraphics[width=1.6in]{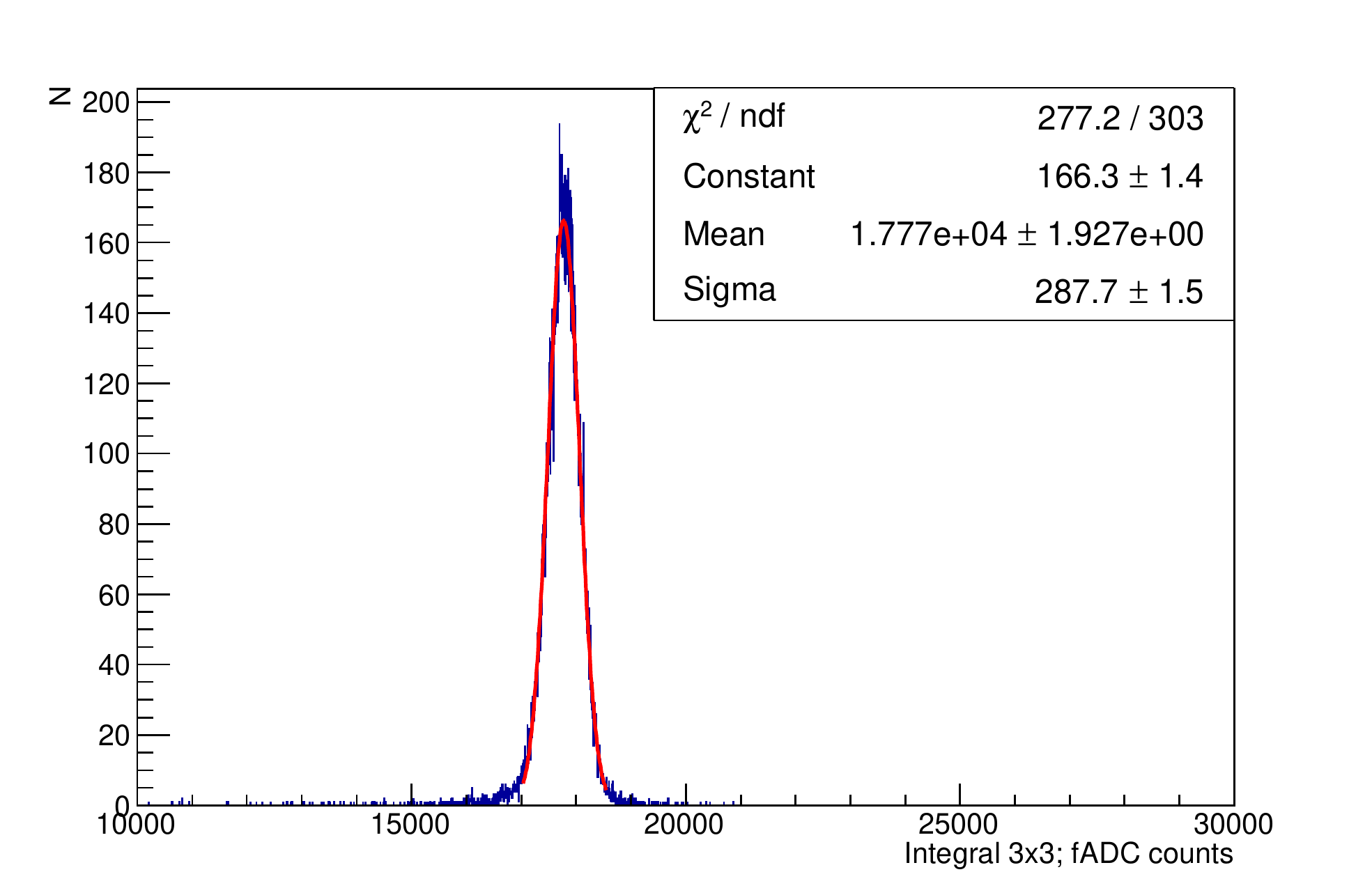}}
\caption{\label{fig:nps-prototype3} (Color online) Total energy reconstructed in the 3x3 calorimeter for 4.7 GeV electrons
}
\end{figure}

\section{Glass scintillators as alternative to crystals}
\label{sec-glass}

Glasses are much simpler and less expensive to produce than crystals and thus offer great potential if competitive performance parameters can be achieved. Early tests have shown good quality and radiation hardness. Due to the different properties, glass would require a 40 cm longitudinal dimension, but could be made to size for different detector regions. 

In the past, production of glass ceramics has been limited to small samples due to difficulties with scale-up while maintaining the needed quality. Some of the most promising materials include cerium doped hafnate glasses and doped and undoped silicate glasses and nanocomposites. All of these have major shortcomings including lack of uniformity and macro defects, as well as limitations in sensitivity to electromagnetic probes. One of the most promising recent efforts is DSB:Ce, a cerium-doped barium silicate glass nanocomposite. Small samples of this material exhibit up to one hundred times the light yield compared to PbWO$_4$ and are in many respects competitive with PbWO$_4$. However, the issues of macro defects, which can become increasingly acute on scale-up, and radiation length still remains to be addressed. 

\section{Summary}
\label{sec-summary}

High resolution electromagnetic calorimeters are an essential piece of equipment at upcoming NPS experiments at 12 GeV Jefferson Lab and the Electron-Ion Collider. This instrument enables precise measurements of DVCS, the method of choice in the program of the three-dimensional imaging of nucleon and nuclei and unveiling the role of 
orbital angular motion of sea quarks and gluons in forming the nucleon spin. 
To satisfy the experimental requirements the EMCal should provide: 
1) good resolution in angle to at least 1 degree to distinguish between 
clusters, 
2) energy resolution to a few \%/$\sqrt{E}$ for measurements of the cluster energy, and
 3) the ability to withstand radiation down to at least 1 degree with respect 
to the beam line. A solution based on PbWO$_4$ would provide the optimal 
combination of resolution and shower width at small angles where the tracking resolution is poor. 

Since the construction of the CMS ECAL and the early construction of the PANDA 
ECAL the global availability of high quality PbWO$_4$ crystals has changed 
dramatically. In this paper we have analyzed samples from SICCAS and samples from CRYTUR, the only two vendors worldwide with mass production capability. Samples were produced between 2014 and 2019. Based on NPS specifications, the overall quality of CRYTUR crystals was found to be better than that of SICCAS samples. Categories in which CRYTUR samples performed better include uniformity of samples, e.g. in transmittance and light yield, and considerably better radiation hardness. CRYTUR samples also showed fewer mechanical defects, both macroscopic and microscopic.

%%% ACKNOWLEDGMENTS %%%
\centerline{ACKNOWLEDGMENTS}

This work is supported in part by NSF grants PHY-1530874, PHY-1306227, PHY-1714133, and the Vitreous State Laboratory (VSL). The detector benefited greatly from components graciously provided by both the PANDA collaboration and Jefferson Lab Hall A/C (crystals and PMTs). We explicitly are grateful to Rainer Novotny and Valera Dormenev from Giessen U. for support with constructing our crystal test setups and providing their facilities for selected optical and irradiation tests.
We would also like to thank Ren-Yuan Zhu from Caltech for tests of selected samples, Craig Woody from Brookhaven National Lab, Steve Lassiter, Mike Fowler and Paulo Medeiros from the Hall C Engineering Group, and Carl Zorn from the Detector Group of the Jefferson Lab Physics Division, for help during various stages of the work.
The Southeastern Universities Research Association operates the Thomas Jefferson National Accelerator Facility under the U.S. Department of Energy contract DEAC05-84ER40150.

%%% BIBLIOGRAPHY %%%

\end{document}